\documentclass[lettersize]{emulateapj}

\usepackage{natbib}
\usepackage{amsmath}
\usepackage{CJK}

\newcommand{\mpc}{\ensuremath{\, h^{-1}\,\mathrm{Mpc} }}
\newcommand{\kpc}{\, h^{-1}\,\mathrm{kpc} }
\newcommand{\kms}{\, \mathrm{km \; s^{-1}}}
\newcommand{\snr}{\ensuremath{\mathrm{S/N}}}
\newcommand{\lya}{Ly$\alpha$}

\newcommand{\waveion}[3]{\ion{#1}{#2} $\lambda$#3}
\newcommand{\wavelya}{Ly$\alpha \; \lambda 1216$}
\newcommand{\wavelyb}{Ly$\beta \; \lambda 1025$}

\newcommand{\beq}{\begin{equation}}
\newcommand{\eeq}{\end{equation}}
\newcommand{\bc}{\begin{center}}
\newcommand{\ec}{\end{center}}
\newcommand{\bfig}{\begin{figure}}
\newcommand{\efig}{\end{figure}}

\newcommand{\fmean}{\ensuremath{\langle F \rangle}}

\newcommand{\taueff}{\ensuremath{\tau_\mathrm{eff}}}

\newcommand{\lambrest}{\ensuremath{\lambda_\mathrm{rest}}}
\newcommand{\lambobs}{\ensuremath{\lambda_\mathrm{obs}}}

\newcommand{\zq}{\ensuremath{z_\mathrm{QSO}}}

\newcommand{\ang}{\ensuremath{\mathrm{\AA}}}
\newcommand{\cpca}{\ensuremath{C_\mathrm{PCA}}}
\newcommand{\cmf}{\ensuremath{C_\mathrm{MF}}}

\newcommand{\dcont}{\ensuremath{\delta C}}
\newcommand{\perpix}{\ensuremath{\mathrm{pixel}^{-1}}}

\def\etal   {et~al.}

\slugcomment{Submitted to AJ}
\shorttitle{MF-PCA Continuum Fitting to SDSS \lya\ Forest}
\shortauthors{Lee \etal}

\begin{document}

\title{Mean-flux Regulated PCA Continuum Fitting of SDSS \lya\ Forest Spectra}
\author{Khee-Gan Lee\altaffilmark{1}, Nao Suzuki\altaffilmark{2}, and David N. Spergel\altaffilmark{1}}
\altaffiltext{1}{Department of Astrophysical Sciences, Princeton University, Princeton, New Jersey 08544, USA}
\altaffiltext{2}{E.O. Lawrence Berkeley National Lab, 1 Cyclotron Rd., Berkeley, CA, 94720, USA}
\email{lee@astro.princeton.edu}

\begin{abstract}
Continuum fitting is an important aspect of \lya\ forest science, since errors in the estimated optical depths
scale with the fractional continuum error. 
However, traditional methods of estimating continua in noisy and moderate-resolution spectra 
($\snr \lesssim 10\; \perpix$ and $R \sim 2000$, e.g. SDSS) 
such as power-law extrapolation or dividing bythe mean spectrum,
 achieve no better than $\sim 15\%$ RMS accuracy. 
 To improve on this, we introduce mean-flux regulated/principal component analysis 
(MF-PCA) continuum fitting. 
 In this technique, PCA fitting is carried out redwards of the quasar
\lya\ line in order to provide a prediction for the shape of the \lya\ forest
continuum. The slope and amplitude of this continuum prediction is then corrected using external 
constraints for the \lya\ forest mean-flux. From tests on mock spectra, we find that 
MF-PCA reduces the errors to 8\% RMS in $\snr \sim 2$ spectra, 
and $<5\%$ RMS in spectra with $\snr \gtrsim 5$. The residual Fourier power in the continuum 
is decreased by a factor of a few in comparison with dividing by the mean continuum, enabling
\lya\ flux power spectrum measurements to be extended to $\sim 2\times$ larger scales. 
Using this new technique, we make available continuum fits for 12,069 $z>2.3$ \lya\ forest spectra from 
SDSS DR7 for use by the community.
This technique is also applicable to future releases of the ongoing BOSS survey, 
which is obtaining spectra for $\sim 150,000$ \lya\ forest spectra at low signal-to-noise ($\snr \sim 2$).
\end{abstract}

\keywords{intergalactic medium --- quasars: emission lines --- 
quasars: absorption lines --- methods: data analysis}

\section{Introduction}
Over the past 2 decades, the Lyman-$\alpha$ (\lya) forest 
absorption observed in the spectrum of
 high-redshift quasars has been an important probe
 of large-scale structure and the inter-galactic medium (IGM) 
 at  $z \gtrsim 2$.
The fundamental quantity of interest of the \lya\ forest is its local optical
depth to absorption, $\tau(\vec{x})$, at position $\vec{x}$.
This is not a directly observed quantity: 
it is derived from the flux transmission $F \equiv e^{-\tau}$, which 
requires knowledge of the intrinsic quasar continuum $C(\lambrest)$
in order to be extracted from the observed flux, $F \equiv f(\lambobs)/C(\lambrest)$, 
where $f(\lambobs)$ is the observed
flux and $\lambrest \equiv \lambobs / (1+\zq) $ is the restframe
wavelength of the quasar at redshift $z = \zq$.
The error in the measured optical depth, $\delta \tau$, scales roughly with the
fractional continuum error, $\delta \tau \sim \delta C / C$, which means that accurate
continuum fitting is important in the optically thin \lya\ forest, where $\tau \lesssim 1$.
Therefore, accurate estimates of the underlying quasar continuum are
crucial in order to take full advantage of modern \lya\ forest data sets. 

While virtually all aspects of \lya\ forest science are dependent on the continuum determination,
some are more sensitive than others.
For example, the  \lya\ flux probability distribution, which is used to constrain the IGM 
temperature-density relation (TDR), is notoriously sensitive to 
continuum errors: \citet{lee:2011} have recently found that a 2\% systematic error in the continuum 
estimation can double the errors in the TDR even with high signal-to-noise data.
On the other hand, measurements of the 1-dimensional flux
power spectrum $P_F(k)$ and other 2-point statistics 
\citep[e.g. threshold clustering functions,][]{lee:2011a}
are affected by the Fourier power introduced by intrinsic quasar emission lines in 
the \lya\ forest region.
For example, the unaccounted continuum variance in the quasar continuum has limited the
 measurement of the 1-dimensional \lya\ forest flux power spectrum \citep{mcdonald:2006}
 to comoving scales of $r \lesssim 40 \mpc$ at $ z = 2.5$. 
 Ongoing attempts to measure the baryon acoustic oscillation (BAO) feature in the
 \lya\ forest using transverse correlations across lines-of-sight are less sensitive
 to continuum errors \citep{mcdonald:2007}; however, large continuum errors will still degrade 
 the significance of the BAO signal measured in this fashion.

Unfortunately, accurate quasar continuum fitting is a non-trivial problem.
At redshifts ($z \gtrsim 2$) in which the \lya\ forest becomes 
accessible to ground-based optical telescopes, the high absorber line-density
makes it challenging to identify the intrinsic quasar continuum. 
With high-resolution and high signal-to-noise (\snr) quasar spectra obtained from large
telescopes, continua are usually fitted using some form of 
spline-fitting to the observed
transmission peaks of the forest --- if one believes that the transmission 
peaks truly reach the quasar continuum at a given redshift \citep[but see][]{fg+08, lee:2011}. 
These direct-fitting methods cannot, in general, be applied to large data sets such as the
Sloan Digital Sky Survey, as
the modest resolution and low \snr\ make it impossible to directly fit
the \lya\ forest except for the very highest \snr\ subsamples ---
and even then steps need to be taken to account for the degradation of
transmission peaks from the lower resolution \citep[see, e.g.,][]{dall+09}.
Moreover, direct fitting techniques usually require significant human intervention and are 
often very time-consuming, precluding their application to the $\sim10^4$ \lya\
forest sightlines in SDSS.

Noisy \lya\ forest data usually require some form of extrapolation from the
relatively unabsorbed spectrum bluewards\footnote{In this paper, 
we use the terms `blue' and `red' relative to the quasar \lya\ emission line unless otherwise noted}
 of the quasar's \wavelya\ broad emission line.
 The simplest way to do this is to fit a power-law $f_{\nu} \propto \nu^{\alpha}$ to spectral
 regions uncontaminated by quasar emission lines redwards of \lya, or 
 even to eschew fitting individual spectra and 
 extrapolate a mean power-law from $\lambrest > 1216\ang$, using 
 power-law values from the literature \citep[from, e.g.,]{vanden-berk:2001}. 
 
 There are two major issues with power-law estimation of \lya\ forest continua.
 Firstly, there is a break in the underlying quasar power-law at $\lambrest \sim 1200\ang - 1300\ang$.
 This was first identified by \citet{zheng:1997} from a study of low-redshift ($\zq \sim 1$) quasars
 observed in the ultra-violet, in which the \lya\ forest continuum could 
 be clearly identified due to the low \lya\ line density at those epochs. 
 A subsequent study by \citet{telfer:2002} found mean power-law indices of 
 $\langle \alpha_\mathrm{NUV} \rangle = -0.69$ and 
 $\langle \alpha_\mathrm{EUV} \rangle = -1.76$ redwards and bluewards of
 $\lambrest \sim 1200\ang$, respectively.
 This implies that a na\"{i}ve power-law extrapolation from $\lambrest > 1216\ang$
 would underestimate the true \lya\ forest continuum by $\sim 10\%$.
Furthermore, \citet{telfer:2002} found a large scatter in $ \alpha_\mathrm{NUV}$
and  $\alpha_\mathrm{EUV}$ from the individual quasars in their sample,
with no correlation between the two; 
this increases the error from power-law extrapolation in individual \lya\ forest spectra.
While \citet{desjacques:2007} and \citet{paris:2011} have discussed 
this EUV-NUV power-law break in the context of \lya\ forest continuum estimation,
it is often ignored in \lya\ forest analyses.

Secondly, the \lya\ forest `continuum' (usually defined around
$\lambrest \approx 1040\ang -1180\ang$) includes weak
emission lines such as \waveion{Fe}{2}{1071} and
\ion{Fe}{2}/\waveion{Fe}{3}{1123}, although the exact identifications vary from 
author to author.
These emission lines can cause deviations of up to $\sim 10\%$ from a flat
continuum. 
It is possible to take these features into account on average: for example, 
\citet{bernardi:2003} modeled them as two Gaussian functions superposed on top 
of an underlying power-law.
However, there is a great diversity in the shape and
equivalent width of these
weak emission lines \citep{suzuki:2006}.
Therefore, the use of an average continuum shape would not account for 
variations of up to 10\% within individual quasars due the presence
of these emission lines.

One possible avenue for improved quasar continuum fits is Principal 
Component Analysis (PCA). \citet{suzuki:2005} explored this using 
a sample of 50 low-redshift quasars observed in the UV by the Hubble Space 
Telescope (HST), in which the $\lambrest < 1216\ang$ continuum can be clearly
identified. They concluded that while PCA fits to the red-side ($\lambrest = 1216-1600\ang$) of
individual spectrum gave a good
prediction of the \lya\ continuum shape (i.e.\ the weak emission lines), 
the overall continuum amplitude had $\sim 10\%$ errors. This is presumably due to 
the EUV-NUV power-law break.
\citet{paris:2011} recently carried out a similar analysis on a high-\snr\ ($\snr \gtrsim 10$ per pixel) subsample 
of the Sloan Digital Sky Survey (SDSS) quasar sample. They found a better 
prediction accuracy of $\sim 5\%$, possibly due to their larger spectral baseline
($\lambrest \approx 1025 \AA-2000 \AA$ as opposed to $\lambrest \approx 1025 \AA-1600 \AA$
in the earlier work).

However, the standard PCA formalism does not take pixel noise into account, whereas 
the majority of quasars observed in SDSS have low signal-to-noise ($\snr < 10$)
\citet{francis:1992} have argued that PCA fitting errors scale directly with the noise level, 
which implies that, e.g., one can expect no better than $\sim 20\%$ continuum 
accuracy in a $\snr=5$ spectrum using PCA, even without taking the power-law break
into account.

For all the reasons outlined above, 
more accurate continuum-fitting methods are sorely needed to take full advantage of 
\lya\ forest data from SDSS and future spectroscopic surveys. In this paper, 
we will explore a refinement of the PCA technique which we term 
`mean-flux regulated PCA' (MF-PCA). 
Briefly, we carry out least-squares-fitting of PCA templates to the unabsorbed
quasar spectrum redwards ($\lambrest >1216\ang$) of the \lya\ line in order
to obtain a prediction for the continuum shape, and then 
use the expected mean flux, $\fmean(z)$, to constrain the amplitude of the 
fitted continuum. 
\citet{tytler:2004} have shown that the dispersion in \fmean\ expected from a 
$\Delta z =0.1$ segment of \lya\ forest at $z=2$ is $\sigma_F(\Delta z = 0.1) \approx 4\%$.
When averaged across an entire \lya\ forest sightline (which spans $\Delta z \approx 0.4$ 
for a quasar  at $\zq=2.5$), one expects the continuum amplitude to be predicted to 
$\sim 2\%$.

This paper is organized as follows: \S~\ref{sec:data} describes the publicly available 
SDSS quasar sample which will be the initial subject of our new technique. \S~\ref{sec:method}
elucidates the MF-PCA technique, which is then tested on mock spectra in 
\S~\ref{sec:mocks}. We will then discuss the results of our continuum fitting and 
future improvements. 
The continuum-fits have been made publicly available and can be downloaded via
anonymous 
 FTP\footnote{\url{ftp.astro.princeton.edu/lee/continua/}}.

\section{Data}\label{sec:data}

The MF-PCA technique which we develop in this paper is optimized
towards large sets of noisy \lya\ forest data spectra. 
We will apply this technique to the Sloan Digital Sky Survey
data, which comprises $\sim 10^4$ \lya\ forest sightlines at moderate
resolution ($R \approx 2000$) and modest signal-to-noise ($\snr \sim \text{few}$ 
per pixel). 

This section provides an overview of the SDSS \lya\ forest data
sample, and also the two sets of quasar templates which we will use to fit this data set.

\subsection{SDSS DR7 \lya\ Forest Sample}

\begin{figure*}

\begin{center}
$\begin{array}{l@{\hspace{0.2in}}l}
\hspace{1em} \normalsize{\textsf{(a)}}  & \hspace{1em}  \normalsize{\textsf{(b)}} \\ [-0.28cm]
\epsfxsize=3.4in
 \epsffile{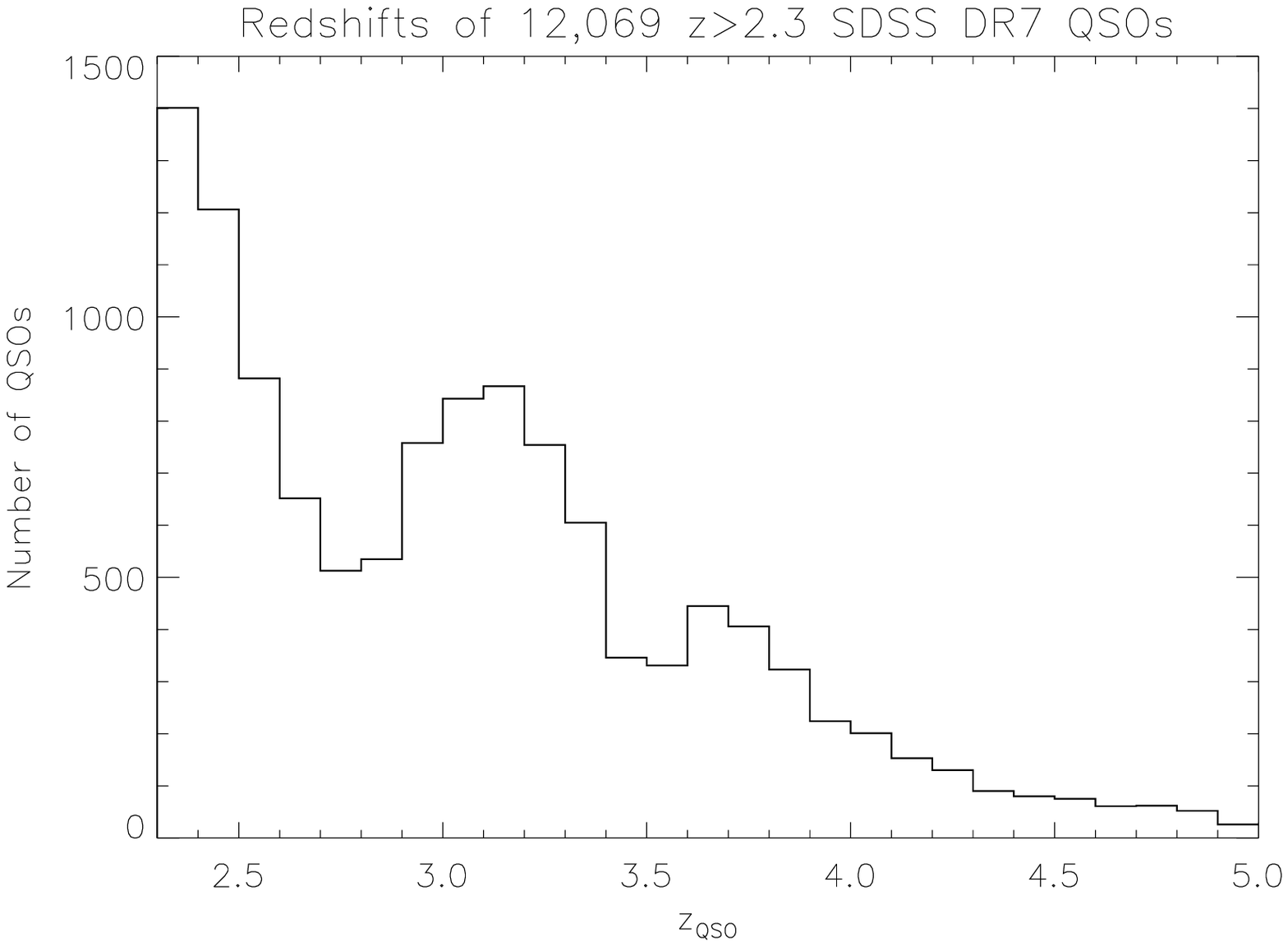} &
	 \epsfxsize=3.4in
 \epsffile{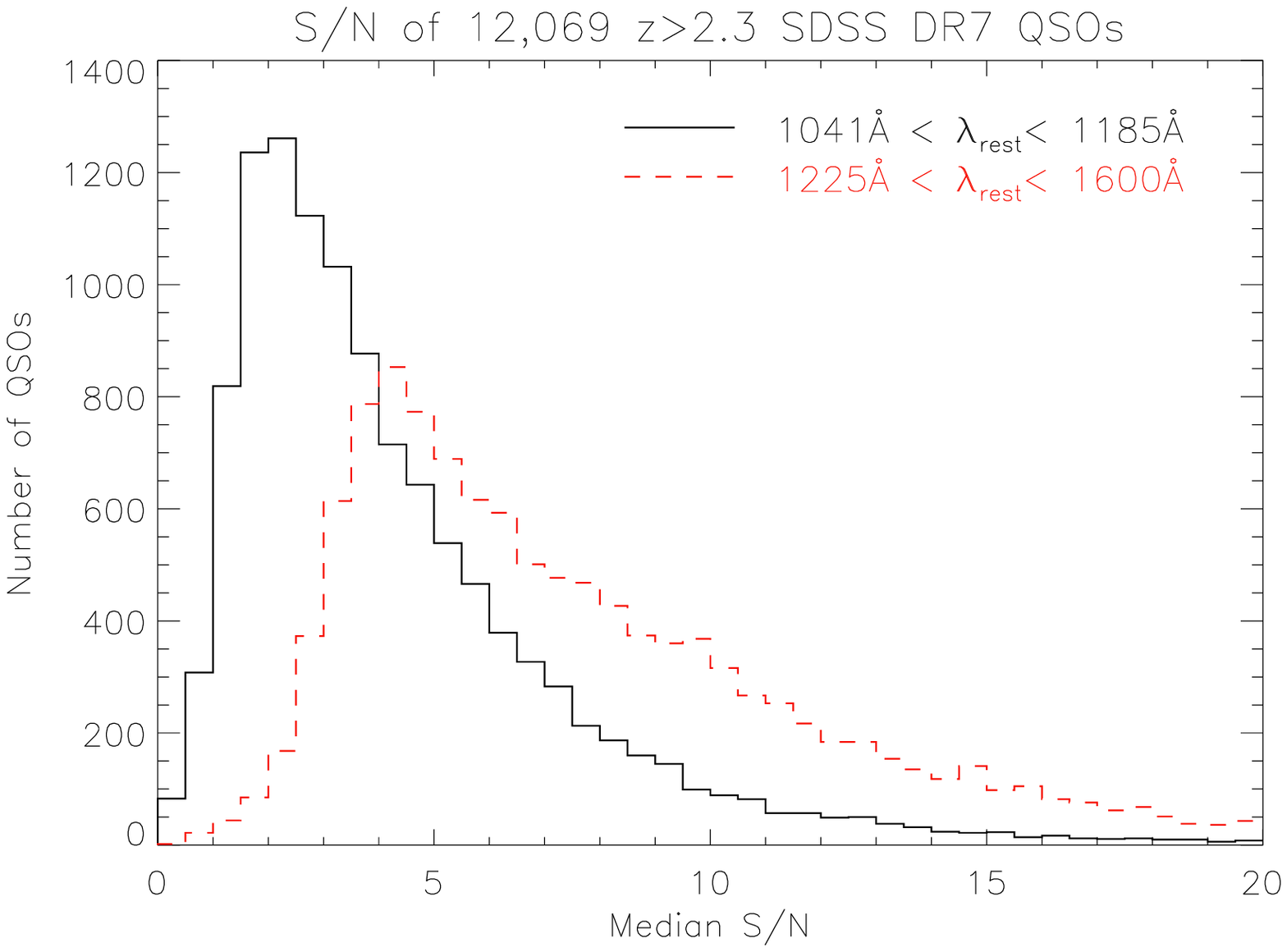}
\end{array}$
\end{center}

\caption{\label{fig:zsnplot}
(a) Redshift distribution of the SDSS DR7 quasars fitted in this paper, in $\Delta \zq = 0.1$ bins. 
We have selected objects
with $\zq \geq 2.3$, which have reasonable coverage of the \lya\ forest. 
53 quasars with $\zq > 5.0$ are not shown in this plot. 
The gaps at $\zq \approx 2.7$ and $\zq \approx 3.5$ are where the quasar colors cross
the stellar locus, making it difficult to select quasars with these redshifts (see, e.g., 
Richards 2002).
(b) Median \snr\ per pixel in our quasar sample, in the \lya\ forest  
($\lambrest = 1041\ang -1185 \ang$; black solid lines) and redwards of 
the quasar \lya\ emission line ( $\lambrest = 1225\ang -1600 \ang$; red dashed lines).
The histograms are in bins of $\Delta \snr = 0.5$.
Note that the majority of the \lya\ forest sightlines have $ \snr < 10 \; \mathrm{pixel}^{-1}$.
}
\end{figure*}

In this paper, we carry out continuum-fitting for publicly-available spectra from 
the final SDSS Data Release 7 (DR7) quasar catalog
\citep{schneider:2010}, which is comprised of 105,783 
spectroscopically confirmed quasars observed from the 2.5m SDSS Telescope in Apache Point, NM.
The spectra cover the observed wavelength range $\lambobs = 3800\ang - 9200\ang$ with a
spectral resolution of $R \equiv \lambda / \Delta \lambda \approx 2000$.

From this overall catalog, we select a subsample suitable for \lya\ forest studies. 
First, we require that some portion of the quasar \lya\ forest region, $\lambrest = 1041 \ang - 1185 \ang$, 
be within the observed wavelength range. Since the extreme blue end (near $\lambobs \approx 3800 \ang$)
of the SDSS spectra are known to suffer from spectrophotometric problems, 
we use $\lambobs=3840\ang$ as the lower wavelength limit.
This sets a minimum quasar redshift of $\zq = 2.3$. 
For the quasars that satisfy this redshift
criterion, we excise the portions of the spectra that lie below $\lambobs=3840\ang$.
In addition, broad absorption line (BAL) quasars have continua which are difficult to characterize
\citep[although see][for a method to recover quasar continua from BALs]{allen:2011}, 
therefore we discard quasars flagged as BALs in the \citet{shen:2011} 
value-added quasar catalog. 

There are 13,133 quasars in the DR7 quasar catalog which satisfy the above criteria. 
We make further quality cuts by discarding 962 spectra which have $\mathrm{SPPIXMASK} = 0-12$ 
bitmasks set \citep[this signifies issues with the fiber; see][for further
details on the SDSS bitmask system]{stoughton:2002}, and 2 spectra
where the signal-to-noise was too low to normalize the spectra, leading to negative normalizations.
This leaves us with a sample of 12,069 spectra, to which we will apply the MF-PCA 
technique.
Within individual spectra, we mask pixels which have either zero inverse-variance or
the $\mathrm{SPPIXMASK} = 16-28$ maskbits set. This avoids the use of
problematic pixels, such as rejected extractions, bright sky-lines or bad flats.

The median signal-to-noise in the sample is $\snr = 3.0$ per $69\,\kms$ SDSS pixel within the \lya\ forest,
and $\snr = 6.2$ per pixel in the $\lambrest = 1225 \ang -1600\ang$ wavelength region.
 The redshift and signal-to-noise\footnote{Henceforth, all signal-to-noise values quoted in this paper are 
the median values per $69 \kms$ pixel, in the range $\lambrest = 1225 \ang -1600\ang$ unless 
indicated otherwise} distributions of our final quasar sample is shown in Figure~\ref{fig:zsnplot}.
It is clear that the \lya\ forest data from DR7 are noisy. Most of the sightlines have median
$\snr < 10$ within the \lya\ forest, which is too noisy to be fitted individually
using existing techniques.

In addition, we need to deal with Damped \lya\ Absorbers (DLAs) within the spectra.
These are absorbing systems with neutral hydrogen column densities of
$N_{HI} \geq 2 \times 10^{20}\; \mathrm{cm^{-2}}$ which result in complete
absorption over large portions ($\Delta v \sim 10^{3} \kms$) of affected sight-lines.
Since the MF-PCA technique (\S~\ref{sec:method}) fits the amplitude of the quasar
continuum based on the mean-flux of the low column-density 
\lya\ forest, the excess absorption of a DLA within a sightline would bias the
continuum estimate.

To correct for this, we use a catalog of 1427 DLAs identified in the SDSS DR7 spectra
by \citet{noterdaeme:2009}. First, we mask the wavelength region corresponding to the 
equivalent width of each DLA \citep{draine:2011}:

\begin{equation}
W \approx \lambda_{\alpha} \left[ \frac{e^2}{m_e c^2} N_{HI} f_{\alpha} \lambda_{\alpha}
\left(\frac{\gamma_{\alpha} \lambda_{\alpha}}{c} \right) \right]^{1/2},
\end{equation} 
where $\lambda_{\alpha}=1216 \ang$ is the rest-frame wavelength of the hydrogen 
\lya\ transition, $e$ is the electron charge, $m_e$ is the electron mass, $c$ is the 
speed of light, $N_{HI}$ is the \ion{H}{1} column density of the DLA, $f_{\alpha}$ is the
\lya\ oscillator strength, and $\gamma_{\alpha}$ is the sum of the Einstein $A$ 
coefficients for the transition.

However, the damping wings of each DLA extend beyond the equivalent width, 
providing a small but non-negligible excess absorption to the pixels close to the
DLA. We correct for this by multiplying each pixel in the spectrum with 
$\exp(\tau_\mathrm{wing}(\Delta \lambda))$, where

\begin{equation}
\tau_\mathrm{wing}(\Delta \lambda) = \frac{e^2}{m_e c^2} \frac{\gamma_{\alpha} \lambda_{\alpha}}
{4 \pi} f_{\alpha} N_{HI} \lambda_{\alpha} \left(\frac{\lambda}{\Delta \lambda}\right)^2 
\end{equation}
and $\Delta \lambda \equiv \lambda - \lambda_{\alpha}$ is the wavelength separation 
in the DLA restframe.

\subsection{Quasar Templates}\label{sec:templates}
 
In order to predict the shape of the \lya\ forest continuum, 
we need a set of template spectra with 
 clearly identified continua at wavelengths $\lambrest < 1216\ang$. 
For this purpose, we will use two different sets of quasar templates, 
derived from quasars observed in the Hubble Space Telescope
 (HST), and SDSS itself.
 
\citet{suzuki:2005} derived PCA templates from 50 quasars that had been observed by
the Far Object Spectrograph (FOS) on the Hubble Space Telescope in the ultraviolet.
At the low-redshifts ($0.14 <\zq < 1.04$) of these quasars, the line-density of the \lya\ forest is 
sufficiently small that the quasar continuum
could be clearly identified. This enabled the creation of templates in the range 
$\lambrest = 1025\ang - 1600\ang$, covering \wavelyb\ to \waveion{C}{4}{1549}.

\citet{paris:2011} recently carried out a similar study, 
applying the techniques in \cite{suzuki:2005} to a subsample of 78 SDSS DR7
quasars. These $\zq \approx 3$ quasars were selected to have full coverage
of the \lya\ forest and relatively high signal-to-noise ($\snr \gtrsim 10 \; \mathrm{pixel}^{-1}$). 
The transmission peaks in the \lya\ forest were
hand-fitted with a low-order spline function to provide a continuum estimate. 
PCA templates were then derived in the spectral range $\lambrest = 1020\ang - 
2000\ang$, which included the \waveion{C}{3}{1906} line.
While this process might give a biased continuum level due to the low
resolution and \snr\ of the templates, it should provide a good description of the
relative shape of the quasar continuum which is required for 
MF-PCA --- the mean-flux regulation process is designed to correct for 
uncertainties in the overall continuum level arising from pure PCA fitting.

We do not expect the redshift differences between the template and the DR7 quasars 
to be a significant issue, even for the \citet{suzuki:2005} quasars ($\langle \zq \rangle \approx 0.6$).
This is because various studies \citep[e.g.][]{vanden-berk:2004, fan:2006}
have suggested that there is little redshift evolution of quasar spectra.
However, the shape of quasar spectra is known to have a significant luminosity dependence, 
such as the well-known Baldwin effect \citep{baldwin:1978}, which is the anti-correlation between the
strength of the \waveion{C}{4}{1549} emission line and the quasar luminosity.
It is therefore reasonable to supposed that there would be a significant difference in the spectral shapes
represented by two templates and the overall SDSS sample. This is because 
the \citet{suzuki:2005} sample is comprised of relatively low-luminosity, nearby quasars as opposed
to the \citet{paris:2011} quasars, which were selected to have high \snr and are therefore a luminous
subsample of the SDSS quasars.

\begin{figure}
\epsscale{1.2}
\plotone{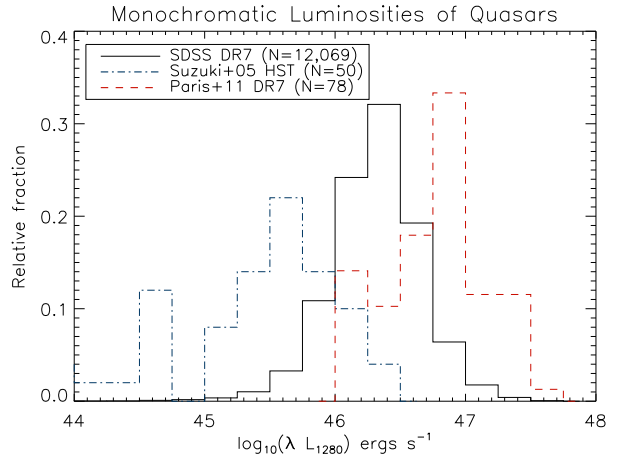}
\caption{\label{fig:lumdist}
Intrinsic luminosity distribution of quasars from SDSS DR7 (12,069 spectra; black solid line),
\citet{suzuki:2005} (50 spectra; blue dot-dashed line), and \citet{paris:2011} (78 spectra; red dashed-line), 
as estimated from $\lambda L_{1280}$. 
This histograms have bin widths of $\Delta \log_{10}(\lambda L_{1280}) = 0.25$, 
and are normalized such that the sum of all the bins in each histogram is unity.
}
\end{figure}

In order to compare the relative luminosities, 
we calculate $\lambda L_{1280}$, the intrinsic monochromatic luminosity
near $\lambrest = 1280 \ang$, for the quasars in our SDSS DR7 
sample as well as the two template samples.
We assume a standard $\Lambda$CDM cosmology with $h=0.7$, $\Omega_m = 0.28$, 
and $\Omega_m + \Omega_{\Lambda} = 1$.
The respective distributions of  $\lambda L_{1280}$ is shown in Figure~\ref{fig:lumdist}. 
The SDSS DR7 quasars have a typical luminosity of $\lambda L_{1280} \approx 10^{46.3} \text{ergs/s}$, 
while the \citet{suzuki:2005} and \citet{paris:2011} quasars are about 0.5 dex fainter and brighter, 
respectively. 
However, the combined luminosity distributions of the \citet{suzuki:2005} and \citet{paris:2011} 
template quasars significantly overlap the full range of SDSS DR7 quasars, which justitifes
the use of both templates in this paper.

\section{Method}\label{sec:method}
The mean-flux regulated PCA (MF-PCA) fitting method described in this 
paper is essentially a two step process: (1) fitting of the 
red side ($\lambrest > 1216\ang$) of the individual quasar spectra 
using PCA templates, to 
predict the shape of the weak emission lines in the \lya\ forest
continuum. This is followed by (2) constraining the amplitude
of the predicted \lya\ forest continuum to be consistent with 
existing measurements of the mean-flux evolution of the 
\lya\ forest, $\fmean(z)$.

\subsection{Least-squares PCA Fitting} \label{sec:pcafit}
The basic concept of principal component analysis (PCA) is
that a normalized quasar spectrum, $f(\lambda)$, 
can be represented as
\beq \label{eq:pca}
f(\lambda) \approx \mu(\lambda) + \sum^{m}_{j=1} c_{j} \xi_j(\lambda),
\eeq
where $\mu(\lambda)$ is the mean quasar spectrum, 
 $\xi_j(\lambda)$ is the $j$th principal component or `eigenspectrum', 
and $c_{j}$ are the weights for an individual quasar.
The formalism for deriving the eigenspectra and weights is 
described in \citet{suzuki:2005} and \citet{paris:2011} . 

The standard PCA formalism for deriving the weights, $c_j$, 
does not take into account spectral noise, which 
renders it unsuitable for noisy SDSS spectra (see Figure~\ref{fig:zsnplot}b).
Instead, we first carry out a least-squares fit to the red-side of each 
spectrum using the full $\lambrest \approx 1000\ang - 1600 \ang$ 
eigenspectra as a basis. Due to the correlation between the weak emission
lines within the \lya\ forest and in $\lambrest \sim 1300\ang - 1500\ang$ \citep{suzuki:2005},
we expect this to provide a reasonable prediction for the shape of the continuum.

\begin{figure*}[t] 
\begin{center}
$\begin{array}{c@{\hspace{0.4in}}c}
\epsfxsize=3.4in
\epsffile{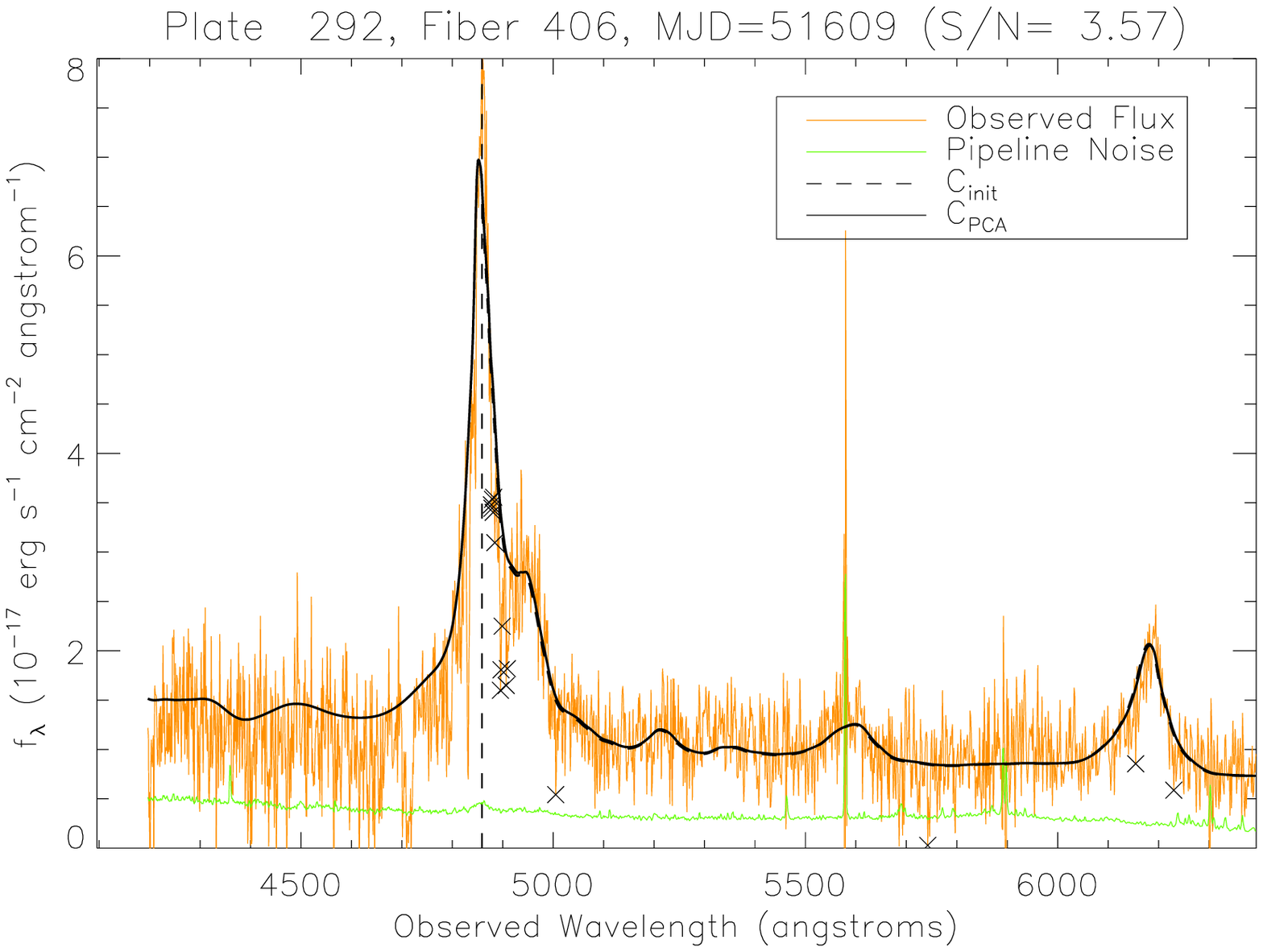} &
	\epsfxsize=3.4in
	\epsffile{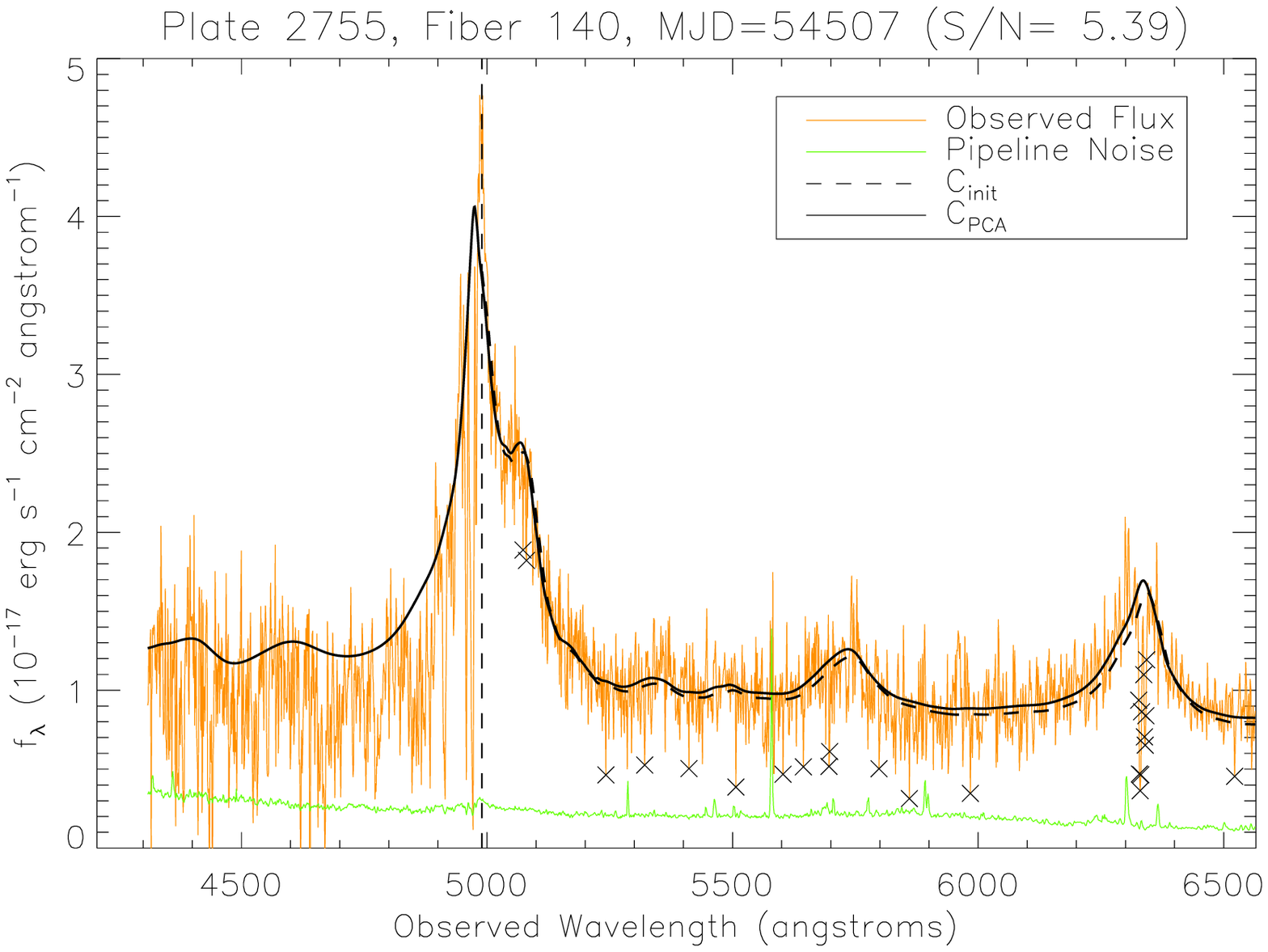} \\ [0.4cm]
\epsfxsize=3.4in
\epsffile{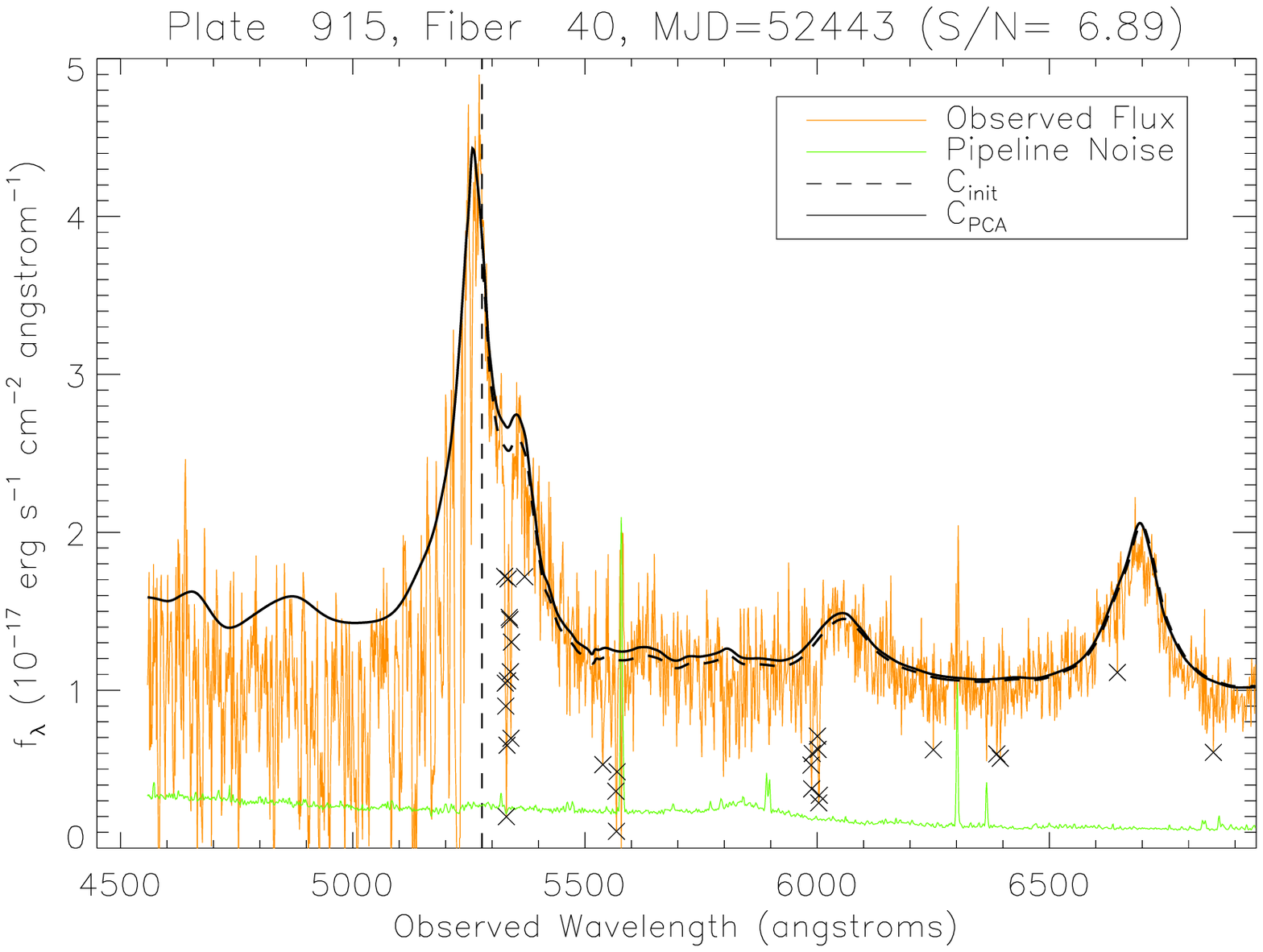} &
	\epsfxsize=3.4in
	\epsffile{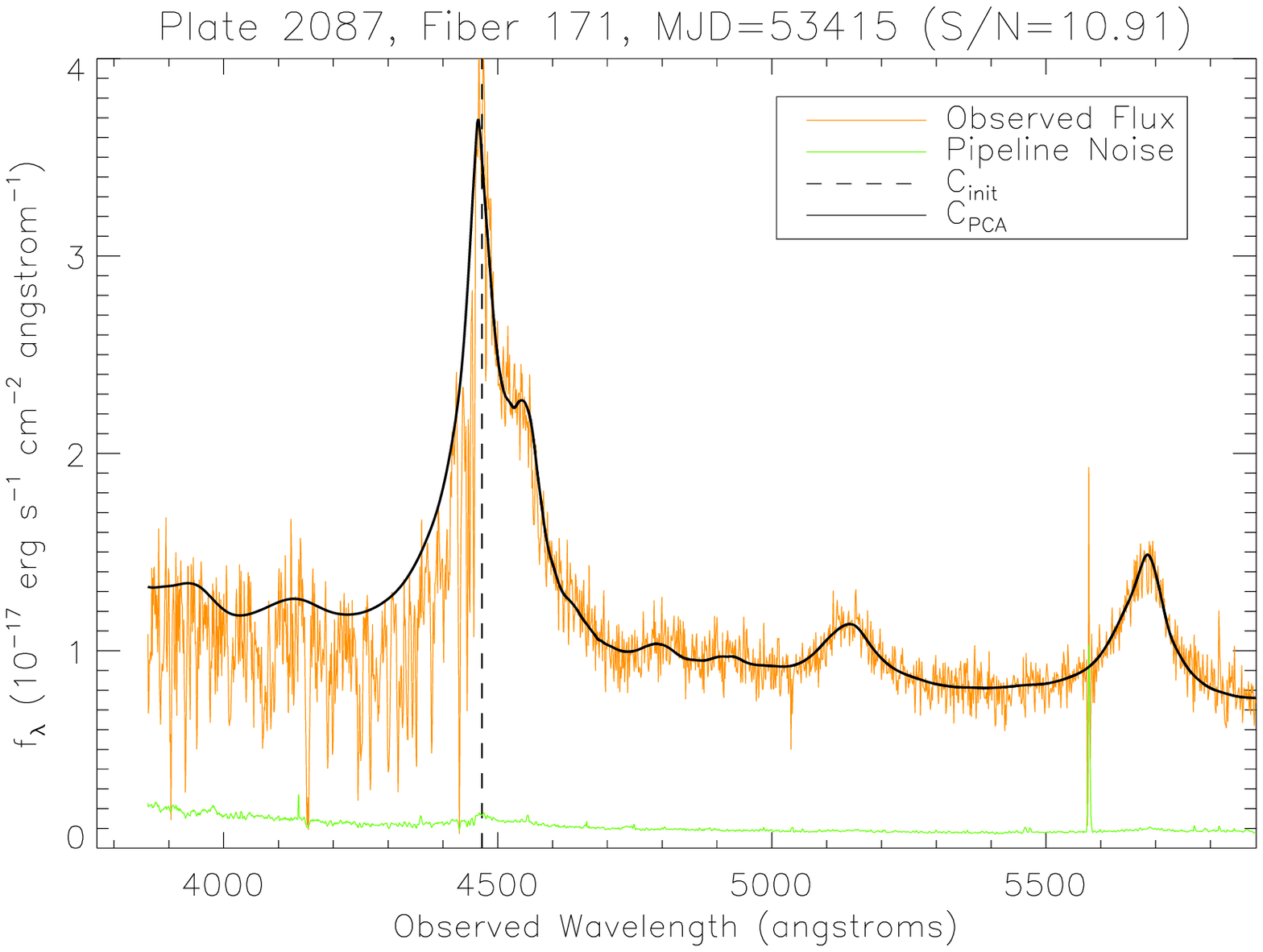} \\ 
\end{array}$
\end{center}
\caption{Successful examples of our least-squares PCA fitting method on
 SDSS quasar spectra with different
\snr. In each plot, we show
the observed flux (orange), pipeline noise (green), $C_{\mathrm{init}}$, 
the PCA fit from the first 
line-masking iteration (black dashed-line), 
and final PCA fit, \cpca (black solid line). Crosses indicate pixels which have been 
discarded by our absorption line-masking scheme. The vertical dashed 
lines indicate $\lambrest = 1216\ang$ in the quasar restframe; 
all fitting is carried out redwards of this wavelength. 
The median \snr\ value quoted is evaluated redwards of the \lya\ emission line, and
the absolute flux error, $|\delta F|$, is defined in Equation~\ref{eq:afluxerr}.
Note that the amplitude of the \lya\ forest continuum ($\lambrest < 1216\ang$) is not well-fitted
by the PCA procedure, and will need to be corrected in the mean-flux regulation
step (\S~\ref{sec:mf})
}
\label{fig:pcafit}
\end{figure*}

As described in \S~\ref{sec:templates}, we have two separate sets of PCA eigenspectra from \citet{suzuki:2005}
and \citet{paris:2011}. In principle, one could combine the two sets of 
quasar templates to generate one set of eigenspectra which would 
encompass the diversity of both template samples. However, the 
template spectra from \citet{paris:2011}  are not available to us at time of writing, 
therefore we will carry out our fitting procedure separately for the two 
sets of PCA eigenspectra. \citet{suzuki:2005} had found that out of their
10 principal component eigenspectra, only the first 8 components appeared to 
describe physical features in the spectra, while the  9th and 10th components seemed
to describe mostly noise. 
Therefore, we will use only 8 components from each set of eigenspectra for our fits. 
In addition, for the sake of consistency 
we limit ourselves to the rest-wavelength range $\lambrest = 1020\ang - 1600\ang$
of each eigenspectrum even though the 
\citet{paris:2011} eigenspectra extend up to $\lambrest = 2000\ang$.

However, 
we have found that fitting the SDSS spectra with just the PCA weights $c_j$
was insufficient to account for the large diversity of the sample.
Therefore, we introduce 2 additional fit parameters: a 
power-law component, $\alpha_{\lambda}$, 
and redshift-correction factor, $c_z$. 
The power-law component, $\alpha_{\lambda}$, 
is necessary due to the large range of slopes found in the SDSS quasars.
Even though the 3rd through 5th principal components in the HST eigenspectra include 
the spectral slope, they also describe some emission-line features --- the 
introduction of $\alpha_{\lambda}$ as a free parameter allows an additional degree of freedom 
and enables a better fit to the emission lines and slope simultaneously.
Due to this degeneracy between the slopes within the eigenspectra and $\alpha_{\lambda}$,
we do not interpret the latter as the slope of the underlying quasar power-law continuum.
The power-law parameter also helps account for low-order spectro-photometric errors 
as well as dust extinction in the spectrum.

The redshift-correction factor, $c_z = \lambrest^{\mathrm{fit}}/\lambrest^{\mathrm{pipe}}$, 
translates the spectrum along the wavelength axis with respect to the rest wavelength given by the
pipeline redshift, $\lambrest^{\mathrm{pipe}}$, 
to a best-fitting rest wavelength, $\lambrest^{\mathrm{fit}}$.
It is required as the SDSS pipeline
redshifts are not completely accurate \citep[see, e.g.,][]{hewett:2010}. 
However, due to the asymmetry and velocity shifting of quasar emission lines at different redshifts, 
we do not necessarily interpret $c_z$ as a true redshift correction ---
it is merely an ad-hoc parameter to obtain the best-possible fit to the spectrum.
The full list of free parameters for our continuum-fitting procedure shown in Table~\ref{tab:params}
($a_\textrm{MF}$ and $b_\textrm{MF}$ are free parameters for the mean-flux regulation step, described
in \S~\ref{sec:mf}).


\begin{table}
\caption{\label{tab:params} Free Parameters in MF-PCA Continuum Fits}
\begin{tabular}{l l}
\tableline
\tableline\\
Fit Parameter     	& Description \\
\tableline \\
$f_{1280}$         	& Flux normalization, evaluated at $\lambrest \approx 1280\ang$  \\
$c_z$                   	& Redshift correction factor \\
$\alpha_{\lambda}$ 	& Power-law exponent \\
$c_1 \cdots c_8$	& PCA coefficients \\
$a_\mathrm{MF}$  	& Linear mean-flux regulation coefficient \\
$b_\mathrm{MF}$	& Quadratic mean-flux regulation coefficient \\
\tableline
\end{tabular}
\end{table}

We are now in a position to carry out the fitting procedure. 
First, the quasar spectrum is shifted to the quasar restframe using the pipeline redshift, 
and normalized at $\lambrest = 1275-1285 \ang$. 
We then use the least-squares fitting routine MPFIT \citep{markwardt:2009} to 
find the best-fitting set of parameters, $[c_z, \alpha_{\lambda}, c_j]$, 
given the spectrum and its noise. 
The initial fits from this procedure is represented by the black dashed lines in 
the examples shown in Figure~\ref{fig:pcafit}.

However, while the intrinsic quasar spectrum is generally well-defined redwards of \lya\ 
in the SDSS spectra, in many cases intervening metal absorption lines 
can be seen in the spectrum in the $\lambrest \approx 1216-1600 \ang$ wavelength. 
To prevent these absorption features from biasing the PCA fitting, we 
carry out a simple iterative procedure to mask these absorption lines:
using the continuum, $C_\mathrm{init}$, obtained from the initial least-squares fit, 
we mask pixels in which $f(\lambda) - C_\mathrm{init}(\lambda) < -2.5\, \sigma(\lambda)$,
where $f(\lambda)$ and $\sigma(\lambda)$ are the observed spectrum and pipeline
noise, respectively. 
We then make a new PCA fit and repeat this process until the fit converges.
In Figure~\ref{fig:pcafit}, the final PCA fits, $C_\mathrm{PCA}$, are shown as black solid
lines while
the masked pixels are denoted by crosses.

From Figure~\ref{fig:pcafit}, we see that the least-squares PCA fitting procedure
generally works well, even with noisy ($\snr \sim \text{few}$) spectra.
The fit to the \wavelya\ emission line is sometimes imperfect, but unsurprising considering
that the fitted range ($\lambrest = 1216 - 1600 \ang$) only takes partial account
of the line. 
Comparing the initial (dashed-line) and final (solid-line) fits in Figure~\ref{fig:pcafit},
 we see that the red-side metal
absorption lines usually have little effect on the fits, but in certain cases
(e.g.\ Figure~\ref{fig:pcafit}(b) and (c)) absorption line-masking
noticeably improves the fit.

\begin{figure*}

\begin{center}
$\begin{array}{l@{\hspace{0.15in}}l}
\hspace{1em} \normalsize{\textsf{(a)}}  & \hspace{1em}  \normalsize{\textsf{(b)}} \\ [-0.28cm]
\epsfxsize=3.5in
 \epsffile{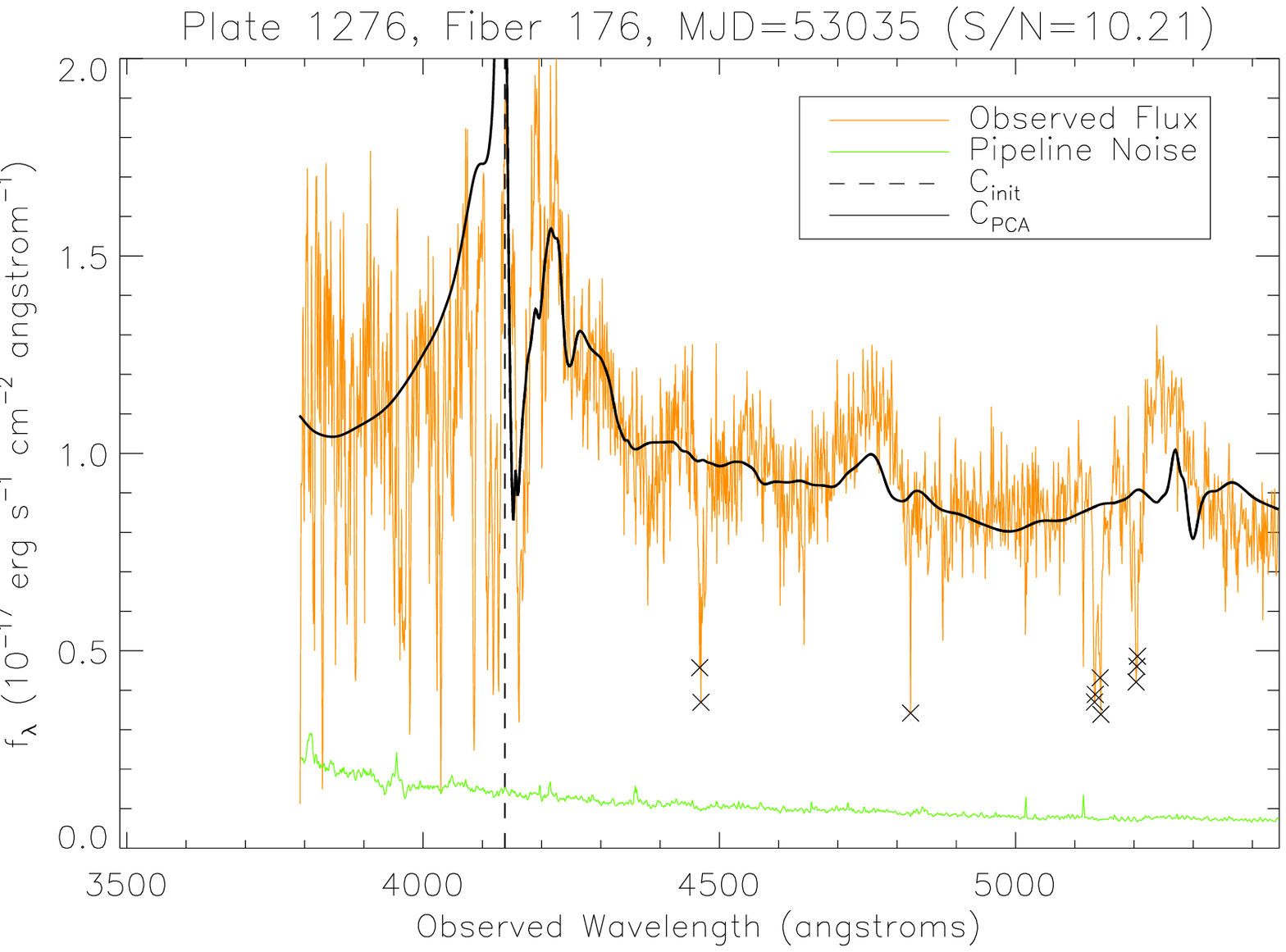} &
	 \epsfxsize=3.5in
 \epsffile{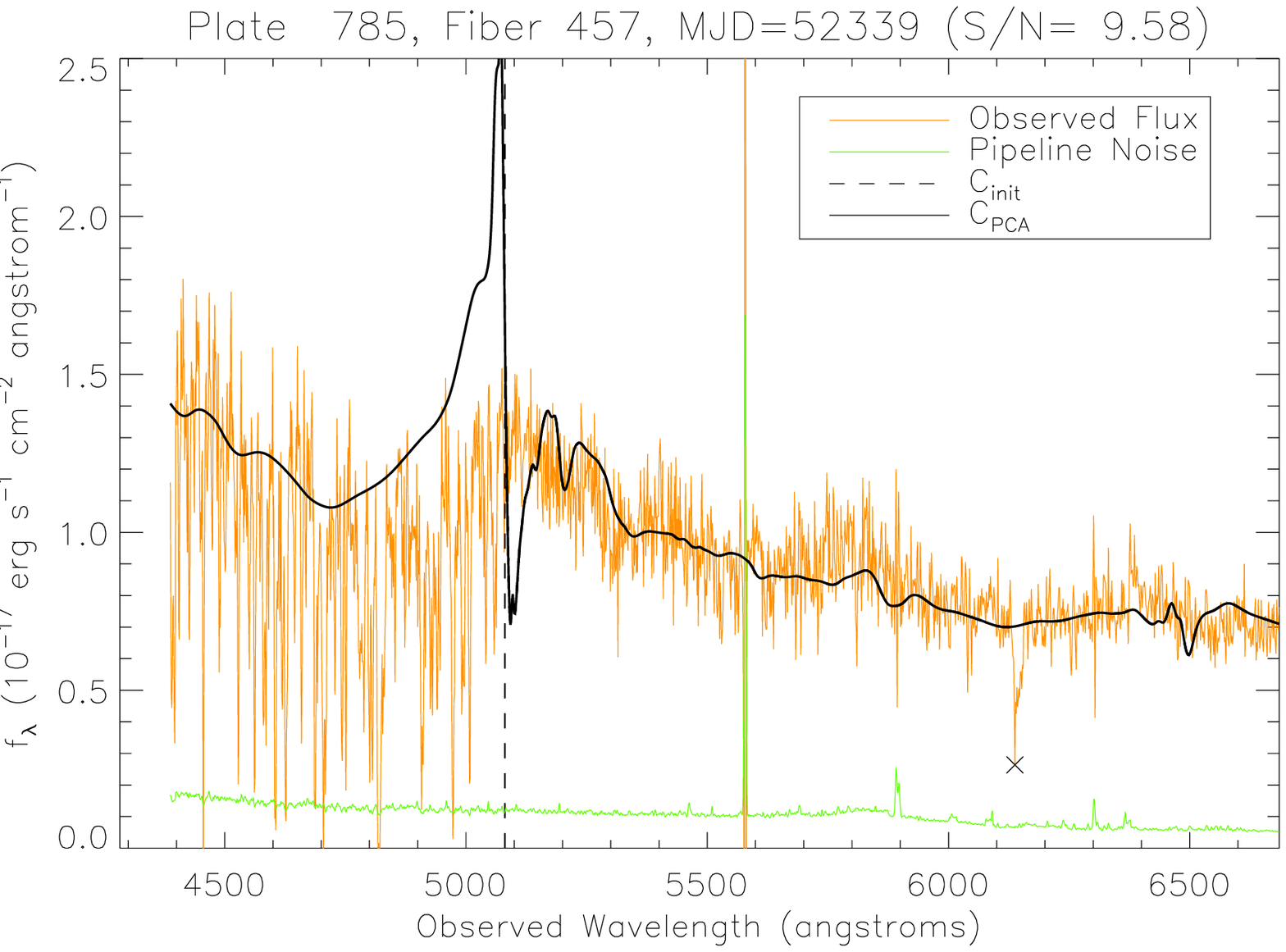}
\end{array}$
\end{center}
\caption{\label{fig:failedfits}
Examples of SDSS quasar spectra in which the PCA fitting procedure fails to provide a reasonable
fit redwards of the quasar \lya\ line. 
(a) A strong proximate \lya\ absorber has decimated the quasar \lya\ emission line, and its associated
\ion{N}{5}+\ion{C}{4} have introduced broad absorption features to the $\lambrest > 1216\ang$ fitting region --- the 
current algorithm is incapable of masking such strong absorption features. 
(b) A weak emission-line quasar. The quasar templates described in \S~\ref{sec:templates}
do not include spectral shapes such as these.
}
\end{figure*}

We use the absolute flux error to quantify the goodness of the PCA fits redwards of \lya:
\beq \label{eq:afluxerr}
|\delta F| = \frac{\int^{\lambda_\mathrm{max}}_{\lambda_\mathrm{min}}
\left| \frac{ \cpca(\lambda) - \tilde{f}(\lambda) }{\tilde{f}(\lambda) }
 \right| d\!\lambda}
{\int^{\lambda_\mathrm{max}}_{\lambda_\mathrm{min}} d\!\lambda },
\eeq
where $\cpca(\lambda)$ is the fitted continuum and $\tilde{f}(\lambda)$ is the 
observed spectrum smoothed by a 15-pixel boxcar, to avoid biasing $|\delta F|$
in noisy spectra. $\lambda_\mathrm{max} = 1600 \ang$ and $\lambda_\mathrm{min}=1225\ang$
represents the range over which we calculate $|\delta F|$.

\begin{figure}
\epsscale{1.2}
\plotone{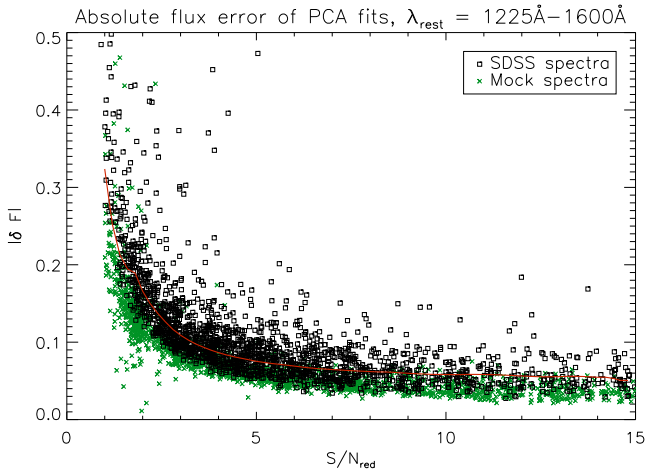}
\caption{ \label{fig:dfred}
Dependence of absolute flux error from the red-side PCA fits, $|\delta F|$, against
the \snr\ per pixel in the range $\lambrest = 1225\ang - 1600\ang$.
This is plotted for random subsets of 2000 SDSS spectra (black squares) and
2000 mock spectra described in \S~\ref{sec:mocks} (green crosses).
The red line traces the 95th percentile of $|\delta F|$ in the mock spectra; 
SDSS spectra with $|\delta F|$ smaller than this are automatically considered good fits,
while spectra above this line are visually inspected to ensure fit quality.
}
\end{figure}

The distribution of $|\delta F|$ in the fitted data is shown in Figure~\ref{fig:dfred}, which plots
$|\delta F|$ against the red-side signal-to-noise per pixel, $\snr_\mathrm{red}$,
for PCA fits to a subset of the SDSS spectra as well as the mock spectra described in
\S~\ref{sec:mocks}. 

Figure~\ref{fig:dfred} provides a useful diagnostic for the quality of 
the PCA fits on the SDSS spectra. 
We expect the fits to the mock spectra (green crosses) to represent 
the case in which the PCA eigenspectra describe the spectra nearly
perfectly (see \S~\ref{sec:mocks}), therefore they 
typically have smaller values of $|\delta F|$ than the real spectra. 
Clearly, the SDSS spectra with large $|\delta F|$ are most likely
bad fits, but note that the presence of metal absorption lines and 
other artifacts in the real data can bias $|\delta F|$ to larger values even for good fits
(we did not mask any lines when calculating $|\delta F|$, as our metal-masking algorithm 
is rudimentary and sometimes masks legitimate pixels). 

In practice, we carry out the PCA fitting procedure using the two different PCA templates
described in \S~\ref{sec:templates}, then for each SDSS spectrum we 
select the fit which gives the lower value of $|\delta F|$. 
Fits with $|\delta F|$ values under the 95th percentile of those from the mock spectra 
(red line in Figure~\ref{fig:dfred} are then automatically considered good fits, 
while the rest are visually inspected and flagged for goodness-of-fit on the red-side of the 
spectrum --- 
approximately 90\% of the spectra were adequately fit.
The spectra which are not well-fitted by our procedure consist mostly of objects
which have strong absorption systems at $\lambrest > 1216\ang$, 
such as metal absorption from DLAs and weak BAL quasars. 
An example of this is shown in Fig.~\ref{fig:failedfits}a.
There are also quasars with unusual spectral shapes which are not represented
in the template spectra described in \S~\ref{sec:templates}, such as quasars
with weak emission lines (Fig.~\ref{fig:failedfits}b).

The spectra now have been had PCA fits carried out on them redwards of \lya, but the
predicted continuum, \cpca, extends bluewards of \lya\ ($\lambrest < 1216\ang$). 
For the objects which are well-fitted on the red-side of the spectrum, 
we expect the predicted continua to provide a reasonable prediction for the shape of the \lya\ 
continuum bluewards of \lya, but the overall amplitude is uncertain due to the EUV-NUV power-law
break described in the Introduction. We now turn to the next fitting
step, mean-flux regulation, to constrain the continuum amplitude.

\subsection{Mean-Flux Regulation}\label{sec:mf}

In the least-squares PCA fitting step described in the previous section, 
we have hitherto used no information bluewards of the quasar
\lya\ emission line due to the absorption from the \lya\ forest. 
However, since the absorption redshift at any point in the \lya\ forest
is known ($z_\mathrm{abs} = \lambobs/1216\ang - 1$), the  
average absorption averaged over each sightline can be used to constrain 
the predicted continuum. 
In this section, we will describe the use of the mean-flux evolution of the \lya\ forest, 
$\fmean(z)$, to regulate the amplitude and slope of the predicted PCA
continuum. 
We refer to this as the `mean-flux regulation' step.

Using the PCA continuum 
$C_\mathrm{PCA} (\lambrest)$ fitted to the observed spectrum, 
we first extract the \lya\ forest transmission 
$F^\mathrm{init}(\lambrest)=f(\lambda)/C_\mathrm{PCA} (\lambrest)$ 
in the range $\lambrest = 1041 \ang -1185 \ang$.
The extracted \lya\ forest is then divided into bins, 
and the mean-flux, $\bar{F}^\mathrm{init}_\mathrm{bin}(\lambda_\mathrm{bin})$, 
is evaluated for each bin, 
where $\lambda_\mathrm{bin}=[1070\ang,1110\ang,1050\ang] $ are the 
central rest wavelengths of each bin.

We now introduce a quadratic fitting function bluewards of a pivot
point, $\lambrest = 1280 \ang$, to obtain the mean-flux regulated
continuum:
\begin{eqnarray} \label{eq:mffunc}
C_\mathrm{MF}(\lambrest) &=& C_\mathrm{PCA}(\lambrest)\nonumber \\ 
& & \times (1 + a_\mathrm{MF} \hat{\lambda}_\mathrm{rest} + 
b_\mathrm{MF} \hat{\lambda}_\mathrm{rest}^2),
\end{eqnarray} 
where $a_\mathrm{MF}$ and $b_\mathrm{MF}$ are free parameters for the fit, 
while $\hat{\lambda}_\mathrm{rest}\equiv \lambrest / 1280\ang - 1$.
Note that for the lower redshifts ($\zq \lesssim 2.4$) in which only a 
limited portion of the \lya\ forest is accessible, we use only
the linear parameter, $a_\mathrm{MF}$, in order to avoid over-fitting.

We again use least-squares-fitting to find the values of 
$a_\mathrm{MF}$ and $b_\mathrm{MF}$ 
which provide the best fit between the extracted mean-flux 
$\bar{F}^\mathrm{fit}_\mathrm{bin}(\lambda_\mathrm{bin})$ and the 
external mean-flux constraint $\langle F \rangle (z)$.
In this mean-flux regulation step, the parameters $c_z, \alpha_{\lambda}$ and $c_j$ fitted
to $\lambrest > 1216\ang$ are kept fixed.
For the mean-flux constraint, we use a double power-law fitted from 
the mean-flux measurement of \citep[][kindly provided by Is\^{a}belle Paris]{paris:2011}:
\beq \label{eq:fmeanz}
 \taueff (z) =
  \begin{cases}
  	\begin{aligned}
   (0.0031 \pm & 0.0012) \times \\
   & (1+z)^{3.49 \pm 0.31},&  z < 3.2
   	\end{aligned} \\
	\begin{aligned}
   (0.0011  \pm & 0.0012) \times \\
   & (1+z)^{4.21 \pm 0.70},&  z \geq 3.2
   	\end{aligned}
  \end{cases},
\eeq
where $\taueff(z) = \langle F \rangle (z)$. 

In principle, the errors in the continuum fit should now be at the level of a few
percent, arising from some combination of the large-scale variance in the \lya\
forest and errors in the fitting. In the next section, we will use mock spectra to quantify the
level of continuum errors in the MF-PCA technique.

\section{Tests on Mock Spectra}\label{sec:mocks}

\begin{figure*}[t] 
\begin{center}
$\begin{array}{c@{\hspace{0.4in}}c}
\epsfxsize=3.4in
\epsffile{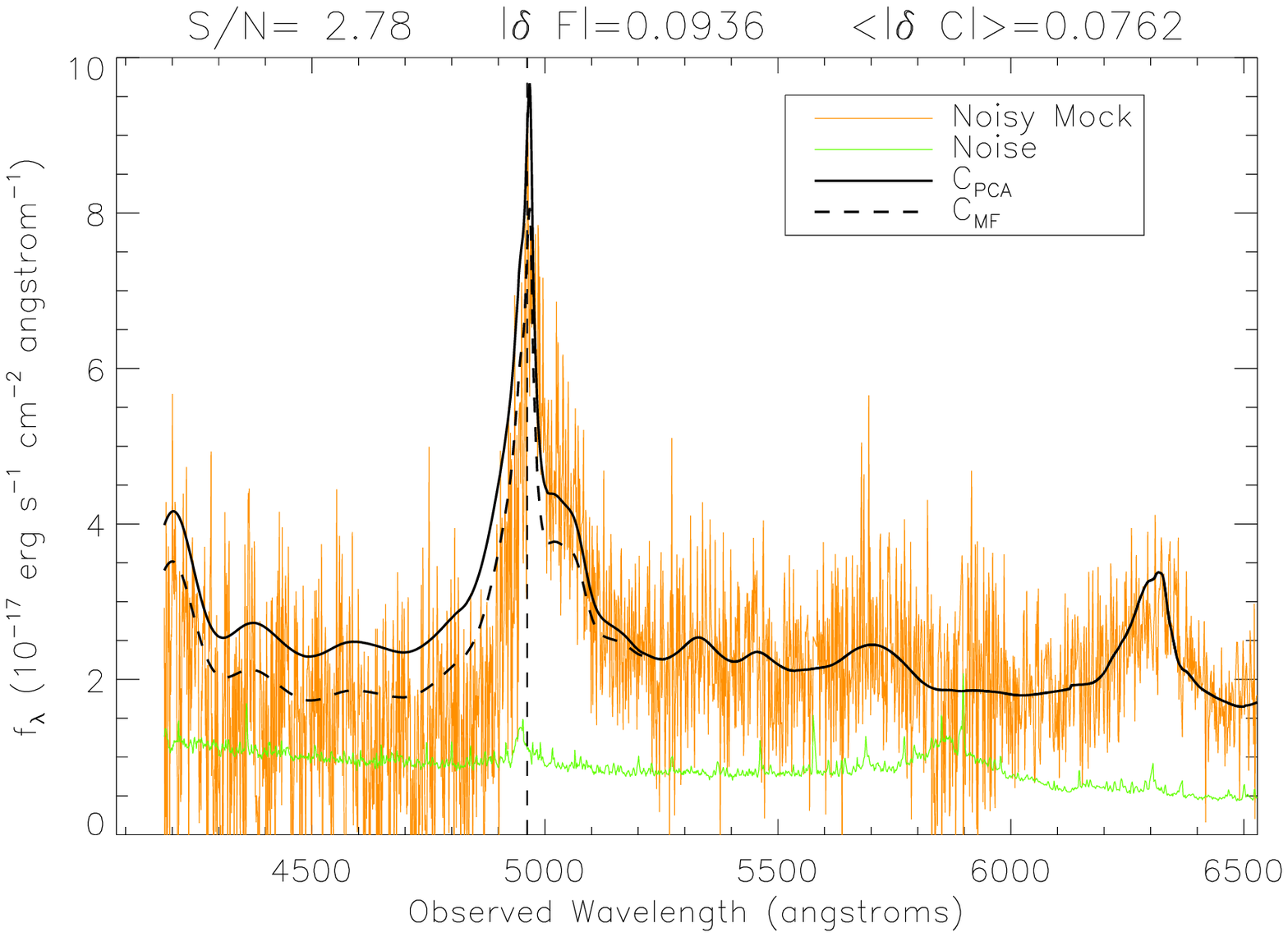} &
	\epsfxsize=3.4in
	\epsffile{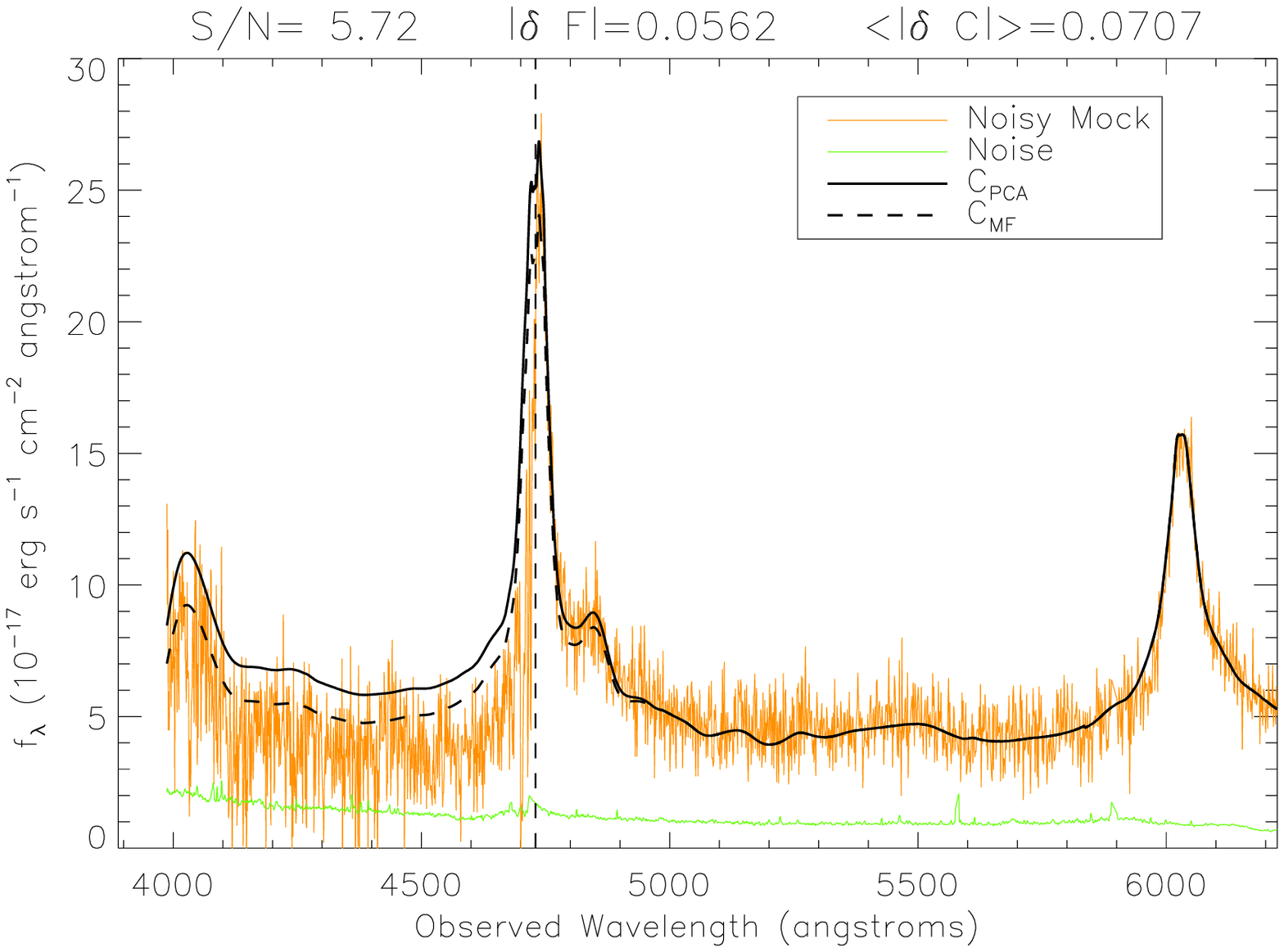} \\ [0.4cm]
\epsfxsize=3.4in
\epsffile{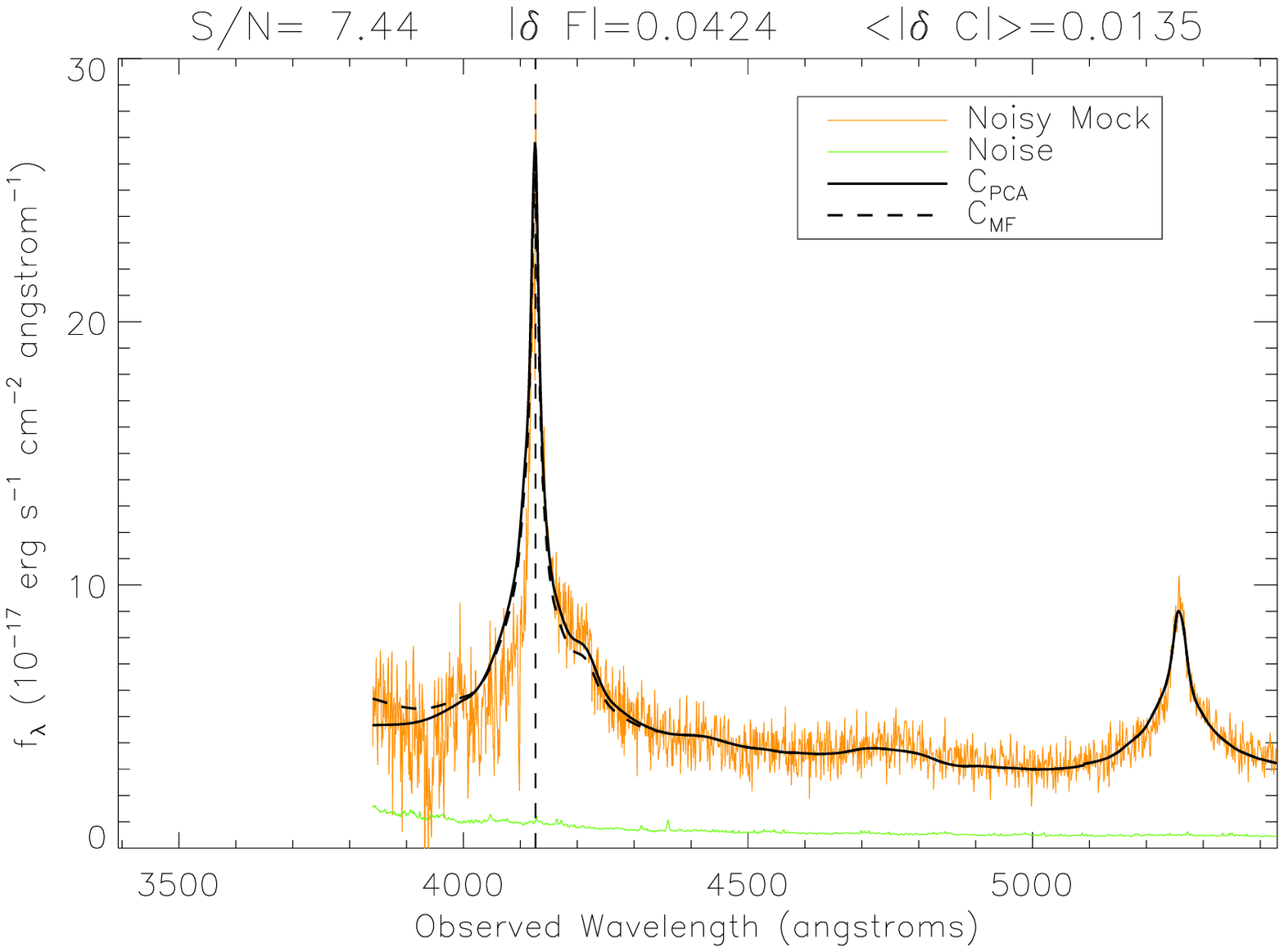} &
	\epsfxsize=3.4in
	\epsffile{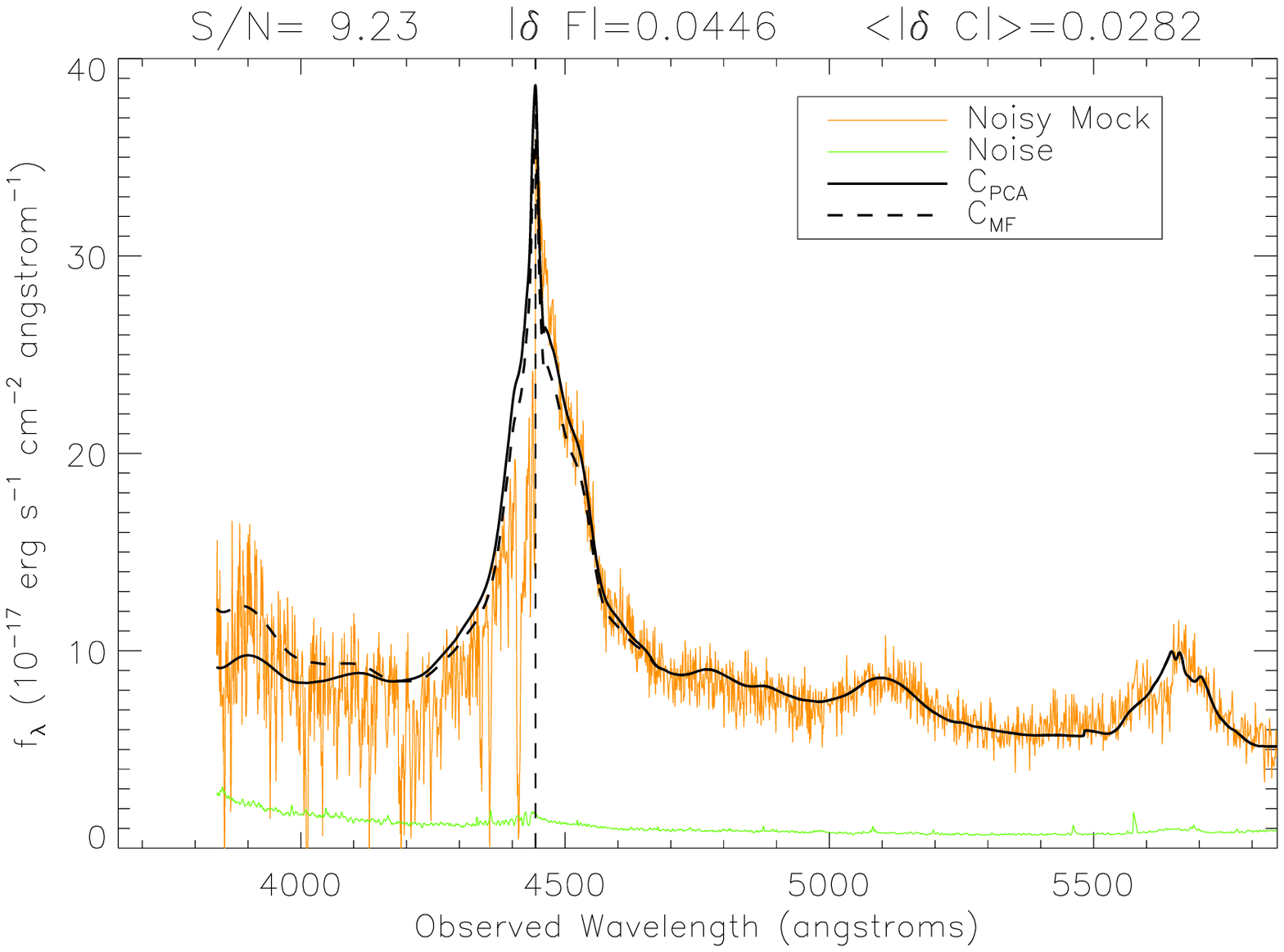} 
\end{array}$
\end{center}
\caption{Tests of the MF-PCA continuum prediction procedure on mock spectra
seeded with noise from real SDSS spectra. In each plot, we show the
noisy mock spectrum (orange); 
pipeline noise (green) used to generate the mock spectrum; \cpca, the least-squares
PCA fit to $\lambrest = 1216\ang -1600\ang$ (black solid line); 
and $C_\textrm{MF}$, mean-flux regulated continuum fit at $\lambrest < 1280\ang$ (black dashed-line). 
The vertical dashed line indicates $\lambrest = 1216\ang$ in the quasar restframe.
$\langle |\delta C| \rangle $ is the RMS continuum-fitting error evaluated over the \lya\ forest
for each individual spectrum.
}
\label{fig:testfit}
\end{figure*}

The errors in the MF-PCA continuum fitting technique can be tested by 
carrying the above procedure on noisy mock spectra, and comparing
the fitted continuum with the `true' continuum which is known by
construction.
This testing process is also useful to check our algorithm
for bugs and efficiency. 
In this section, we will describe the process of generating realistic mock
spectra, and the quantitative results of the MF-PCA technique.
We will also make a
comparison between MF-PCA and the common technique of using the 
mean quasar continuum as the \lya\ forest continum.

\subsection{Generating Mock Spectra}\label{sec:mockgen}

The first step is to create synthetic quasar spectra from PCA
eigenspectra, by making Gaussian realizations of the PCA weights, 
$c_j$, (see Equation~\ref{eq:pca}) in the manner described in \citet{suzuki:2006}. 
Note that this is an approximation, as the distribution of the weights may not be
fully Gaussian, but it does generate realistic-looking quasar spectra.
In principle, the PCA eigenspectra used to generate the mock spectra and
those used in the fitting procedure should be separate but drawn from 
the same distribution. 
While we do have two sets of PCA eigenspectra (\S~\ref{sec:templates}), 
they represent quasars with different luminosities. 
Hence, we do not expect to be able use eigenspectra from \citet{paris:2011}  
to fit mock spectra generated from the
\citet{suzuki:2005} eigenspectra, and vice versa.
However, since we only use 8 principal components in our PCA fitting
step (\S~\ref{sec:pcafit}), we can generate our mock spectra using
10 principal components in order to increase the uncertainty in the fitting.
Nevertheless, the tests described in this section will primarily apply to
the limit in which the PCA eigenspectra are a good representation of
the fitted quasars, which we have argued (\S~\ref{sec:templates}) is a
reasonable assumption.

Therefore, we generate mock quasar spectra in the spectral range $\lambrest = 1020\ang-1600\ang$, 
using 10 principal components from the \citet{suzuki:2005} eigenspectra.
Next, we need to introduce \lya\ forest absorption to the mock spectra. 
For this, we use the publicly available Roadrunner \lya\ forest 
simulations\footnote{\url{http://mwhite.berkeley.edu/BOSS/LyA/RoadRunner/}} of \citet{white+10}.
These are N-body simulations with a box size of
$(750 \mpc)^3$ and a grid scale of $187.5 \kpc$, in which the \lya\ forest flux
was derived using the fluctuating Gunn-Peterson approximation. 
The simulations were released in the form of 22,500 \lya\ forest sightlines per box,
output at redshifts $z_\mathrm{box} \approx 2.00, 2.25, 2.50, \; \text{and}\; 2.75$.

For a given mock quasar at redshift \zq, we select the simulation box with the closest redshift,
$z_\mathrm{box}$.
Using Equation~\ref{eq:fmeanz}, we then re-normalize the mean-flux of 
the box to $\fmean(z=(1+\zq) 1100/1216 - 1)$,
 i.e.\ using the absorber redshift corresponding
to the $\lambrest = 1100 \ang$ in the quasar spectrum.
We choose to normalize the mean-flux across the entire box rather than in individual
spectra in order to preserve the variance across different lines-of-sight, which is a 
source of error in the MF-PCA continuum fitting.
A random line-of-sight is selected from the set of skewers, and the transmitted flux
in each pixel, $F_i$, is rescaled to 
$F'_i = F_i \times \fmean(z_{\mathrm{abs},i}) / \fmean (z=(1+\zq) 1100/1216 - 1)$, where
$z_{\mathrm{abs},i}$ is the absorber redshift corresponding to the pixel. 
This introduces redshift evolution of the mean flux, $\fmean(z)$, within the 
individual sightlines which had hitherto had a fixed value of \fmean.

The simulated \lya\ forest absorption is added to the mock quasar spectrum in $\lambrest < 1216\ang$, 
and smoothed to the approximate SDSS resolution, $R=2000$. 
Gaussian noise is then added to the mock spectrum using the noise array of 
a randomly-chosen SDSS quasar spectrum with the same \zq and \snr.
The mock spectra are then run through the MF-PCA fitting process described
above to obtain continuum fits.

\subsection{MF-PCA Continuum-Fitting on Mock Spectra}
In Figure~\ref{fig:testfit}, we show several examples of the mock spectra and the
fitted MF-PCA continua. The first thing to note is that mock spectra look 
realistic. They look similar to the real spectra shown in Figure~\ref{fig:pcafit},
 apart from the lack of metal absorption redwards of the \lya\ emission line.

As described in \S~\ref{sec:pcafit}, we use the mock spectra as a benchmark for 
the PCA fit quality on the red-side ($\lambrest > 1216\ang$) of the spectra ---
we automatically accept all fits with absolute flux error, $|\delta F|$, less than 
the 95th percentile of the $|\delta F|$ distribution measured from the mocks (Figure~\ref{fig:dfred}).
For the fits with larger $|\delta F|$ that require visual inspection, we use the mocks as a visual guide 
for what constitutes a good fit.

It is clear from Figure~\ref{fig:testfit} that the mean-flux regulated continuum, \cmf, 
is a corrected version of the PCA fit, \cpca. In several cases, the initial PCA
continuum,\cpca, appeared unphysical (e.g.\ dipping below the 
peaks of the forest). These were rectified by the mean-flux regulated fit,
\cmf.

We can place this on a more quantitative footing by comparing the fitted continua, $C_\mathrm{fit}$, 
to the
`true' continua, $C_\mathrm{true}$, which is known by construction in the mock spectra. We define the 
continuum fitting residual, 
\beq
\delta C(\lambrest) \equiv \frac{C_\mathrm{fit}(\lambrest)}{C_\mathrm{true}(\lambrest)} - 1.
\eeq

\begin{figure*}

\begin{center}
$\begin{array}{l@{\hspace{0.15in}}l}
\hspace{1em} \normalsize{\textsf{(a)}}  & \hspace{1em}  \normalsize{\textsf{(b)}} \\ [-0.28cm]
\epsfxsize=3.5in
 \epsffile{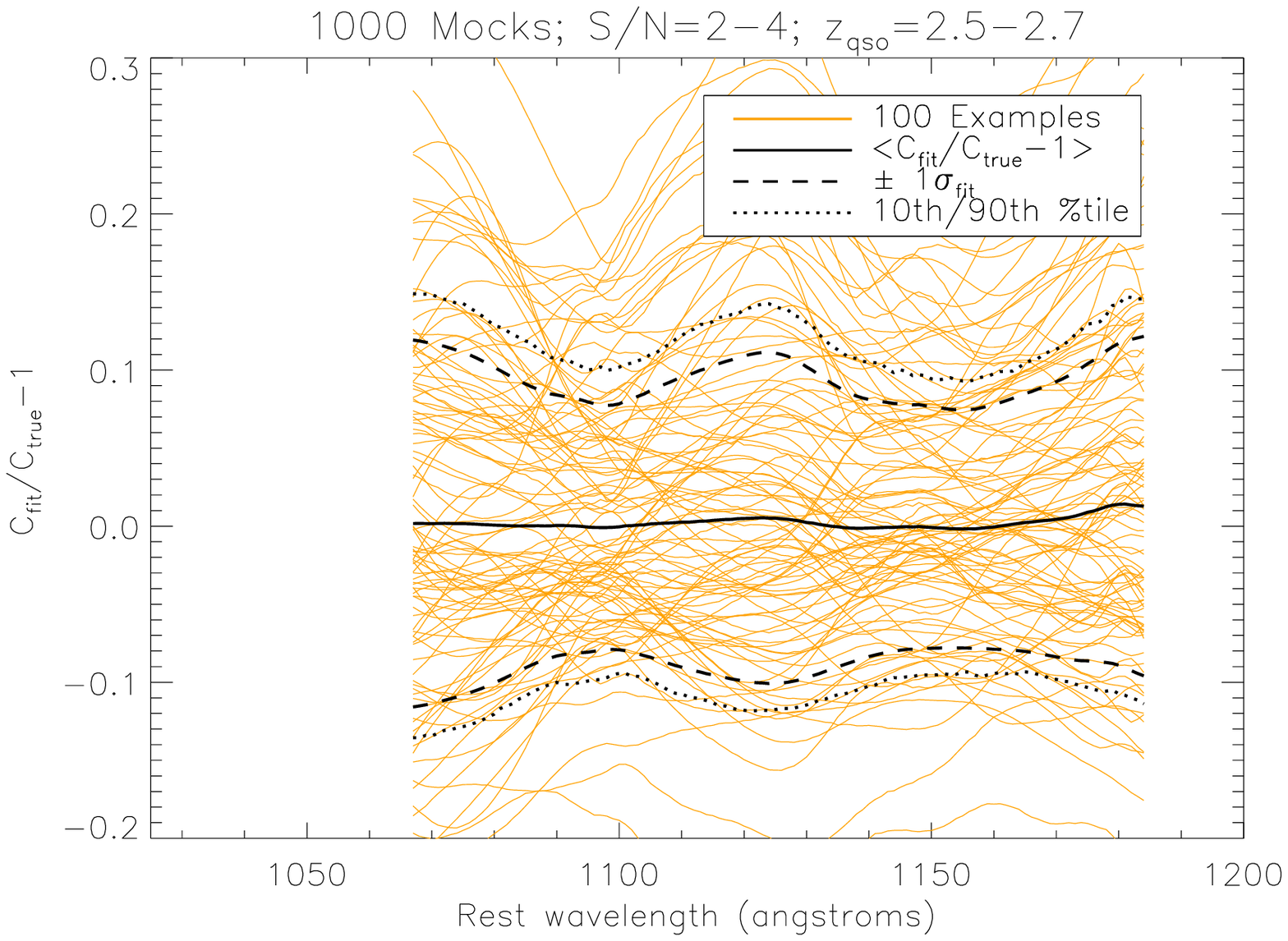} &
	 \epsfxsize=3.5in
 \epsffile{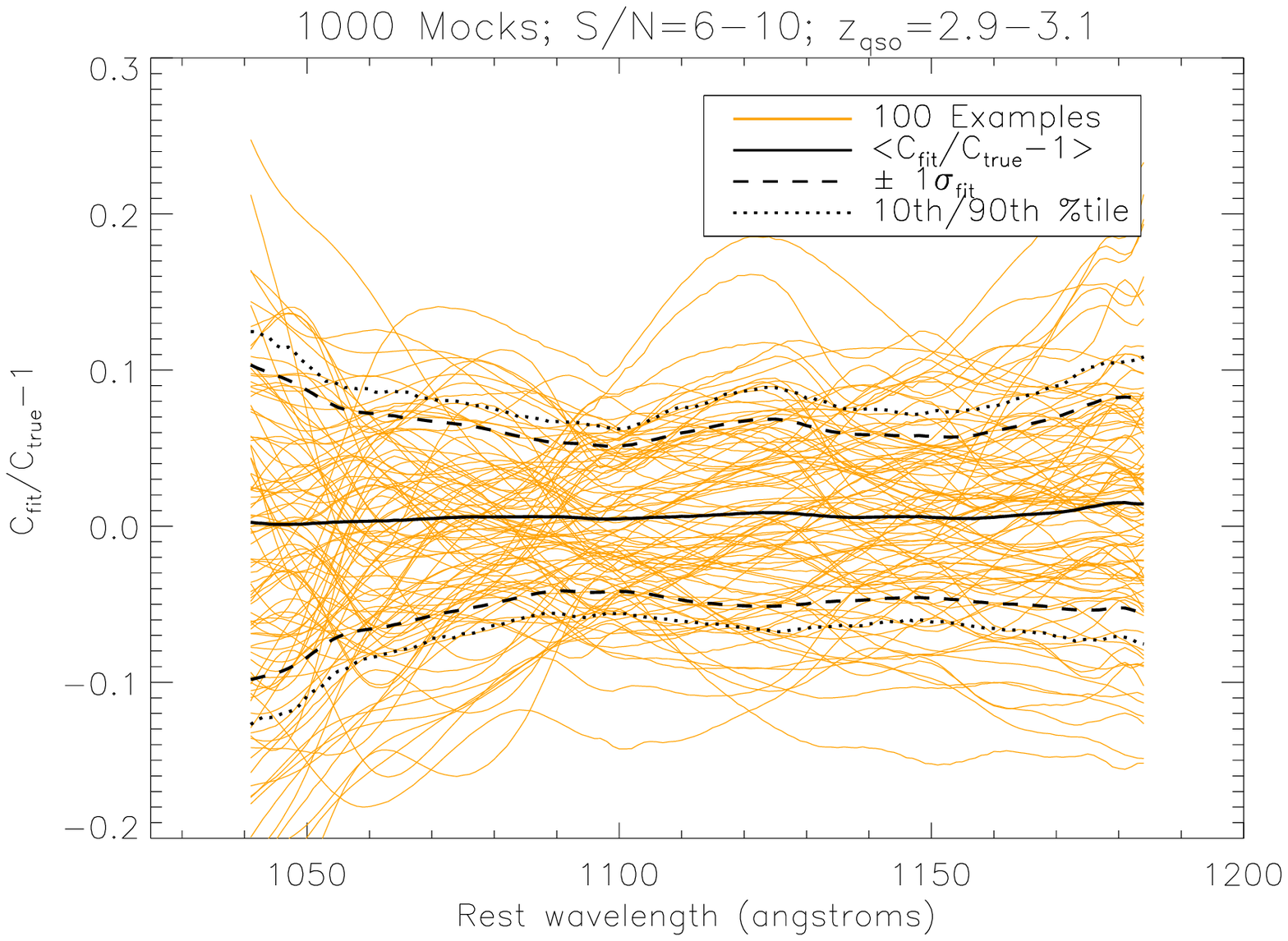}
\end{array}$
\end{center}
\caption{\label{fig:residplot}
Continuum fitting errors from MF-PCA fitting on mock \lya\ forest spectra, plotted as a 
function of quasar restframe wavelength for (a) $2 \leq \snr < 4$ and $2.5 \leq \zq \leq 2.7$; 
(b) $6 \leq \snr < 10$ and $2.9 \leq \zq \leq 3.1$.
In both cases, 1000 mock spectra were generated and continuum-fitted. The grey lines represent
a random subset of continuum-fitting errors from 100 fits. We also show the overall
bias (solid line), the dispersion (dashed line), and the 10th/90th percentiles (dotted line)
of the errors as a function of wavelength. Note that Figure~(a) represents a regime with the worst
MF-PCA continua, and 
is truncated at the lower-limit of the observed spectral range, $\lambobs = 3840\ang$.
}
\end{figure*}

We can then carry out MF-PCA fits on large numbers of mock spectra to obtain statistics 
on the continuum fitting errors as a function of \snr\ and quasar redshift.
In Figure~\ref{fig:residplot} we show the residuals from fitting 1000 mock spectra in two 
bins of signal-to-noise, \snr, and quasar redshift, \zq.
The orange lines show the residuals as a function of wavelength, $\dcont(\lambrest)$, 
binned into restframe $1\ang$ bins
from a subset of 100 mocks. The dashed lines show the $1-\sigma$ dispersion
of the residuals estimated from bootstrap resampling at each $1\ang$ wavelength bin, while
the dotted lines show the 10th and 90th percentile at each wavelength.
Figure~\ref{fig:residplot}(a) represents one of the worst case scenarios
--- the signal-to-noise ($\snr =2-4$) is low at $\lambrest > 1216\ang$, making it difficult
to obtain a good fit for the continuum shape. At the same time,  
the redshift is sufficiently low that the \lya\ forest occupies the blue end $\lambobs \lesssim 4000\ang$
of the SDSS spectrographs where the signal-to-noise deteriorates rapidly with decreasing wavelength.
This causes a large scatter in the flux of \lya\ forest 
even when averaged over large segments, introducing more errors to the mean flux regulation
process.

At moderate signal-to-noise ($\snr = 6-10$, Figure~\ref{fig:residplot}(b)), 
the situation is significantly better. The $1-\sigma$ dispersion of the fit residuals are
well under 10\%, and $6-7\%$ in the central portion of the fitted region. 
Many of the residuals are flat to within a few percent across the \lya\ forest region,
indicating that the PCA-fitting has successfully accounted for the shape of the \lya\ 
continuum. About one-tenth of the fitted continua are badly fit, with badly-predicted continuum 
shapes and/or residuals greater than 10\%. 

We also calculate the mean bias, $\overline{\delta C}(\lambrest)$, 
 of the residuals in Figure~\ref{fig:residplot}.
Averaged over $\sim 10^3$ spectra, the MF-PCA technique have a low 
bias of $< 1\%$, although this not include any systematics errors in the $\fmean(z)$ measurement
used to constrain the overall continuum levels. 

\begin{figure}
\epsscale{1.2}
\plotone{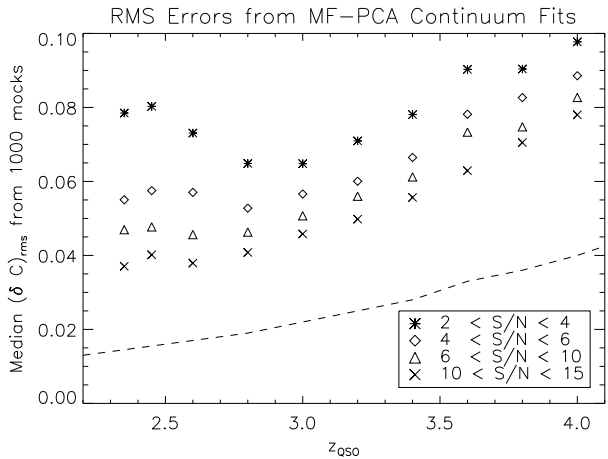}
\caption{ \label{fig:dcont_z}
Median RMS continuum fitting error, $\langle |\delta C| \rangle$, from fitting 1000 mock spectra, 
calculated for different redshifts, \zq, and signal-to-noise, \snr, on the red-side of the spectrum.
The rise in the low-\snr\ values of $\langle |\delta C| \rangle$ at low-$\zq$ is due to the
increase noise levels at the blue end ($\lambobs \lesssim 4000\ang$) of the SDSS spectra, 
while the overall increase in $\langle |\delta C| \rangle$  with redshift is due to the increase in 
the variance of the \lya\ forest. The dashed-line shows the RMS fitting error in the absence of
continuum structure and noise.
}
\end{figure}

To quantify the overall fit quality on each mock spectrum, 
we use the RMS of the continuum residuals
evaluated over $\lambrest = 1041 - 1185 \ang$:
\beq
 (\delta C)_\mathrm{rms} \equiv 
\left[ \frac{\int \left( \frac{C_\mathrm{fit}(\lambrest)}{C_\mathrm{true}(\lambrest)} - 1 \right)^2 d\!\lambrest}
{\int d\!\lambrest} \right]^{1/2}.
\eeq 
Figure~\ref{fig:dcont_z} shows the median RMS error from runs of 
1000 mocks as a function of redshift, for 4 different \snr\ bins. 
At lower redshifts, the RMS error is relatively high for the low-\snr\ spectra 
because the mean flux regulation is affected by 
the increased noise levels near the blue-end of the SDSS spectra at $\lambobs \sim 4000\ang$. 
As the observed \lya\ forest region clears the blue end of the spectra, 
the RMS error decreases to a minimum at $\zq \approx 3.0$.
It then rises with redshift at $\zq > 3$ due to the increasing variance in the \lya\ 
forest, which adds to the error in the mean-flux regulation.
At fixed redshift, the median RMS decreases with \snr\ as might be expected. 
Below $\zq \approx 3$, it drops below $5\%$ RMS for moderate signal-to-noise
($\snr \sim 5 $) and asymptotes
to $\sim 3-4\%$ RMS for the $\snr > 10 $ spectra.

To estimate the contribution from various sources of error in the MF-PCA fitting, we also 
carried out the mean-flux
regulation directly on the simulated \lya\ forest skewers without 
introducing a quasar continuum, or adding pixel noise. In other words, we directly fit the function
Eq.~\ref{eq:mffunc} to the mean flux evaluated over 3 bins in each skewer.
The RMS error in the continuum from this estimate is shown as the dashed line in Fig.~\ref{fig:dcont_z}.
For $\zq = 2.3$ quasars, the RMS error contribution from \lya\ forest variance is about 1.5\% ;
this increases to 4\% at $\zq=4$. This suggests that even in the limit of high-\snr, errors in the continuum
shape from PCA fitting contributes $1-2 \%$ to the overall RMS continuum error.

\subsection{Power-law+Mean Continuum Fitting}

To place the above results in context, we carry out another set of continuum fits on the
mock spectra, using the continuum model 
\beq \label{eq:c_pl}
C_\textrm{mean}(\lambrest) = 
f_{1280}\times  \mu(\lambrest)\times \left(\frac{\lambrest}{1280\ang}\right)^{-\alpha},
\eeq
in which the mean spectrum, $\mu(\lambrest)$, 
is multiplied with a power-law, $\lambrest \propto \lambrest^{-\alpha}$, and $f_{1280}$ is the flux
normalization at $\lambrest=1280\ang$. Both $f_{1280}$ and
$\alpha$ are determined separately for each quasar, with the power-law fitted 
to the regions near $\lambrest = 1280\ang$ 
and $\lambrest = 1450\ang$. This model is highly similar to that implemented in \citet{slosar:2011}.

\begin{figure}
\epsscale{1.2}
\plotone{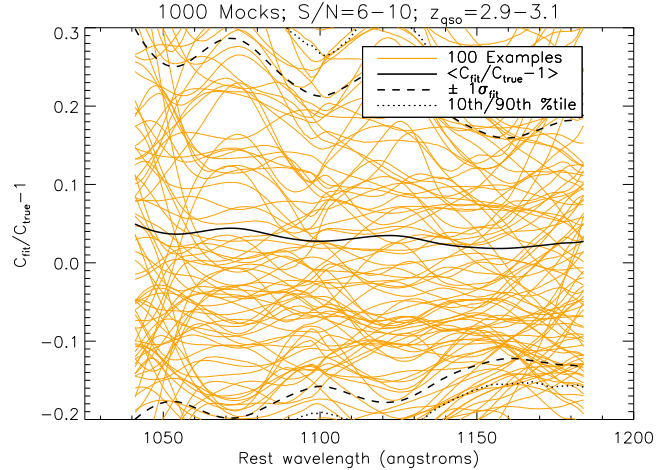}
\caption{ \label{fig:residplot_plaw}
Same as Figure~\ref{fig:residplot}(b), but for the power-law+mean spectrum continua, 
$C_\textrm{mean}$ (Equation~\ref{eq:c_pl}), fit to 1000 mock spectra. Note the larger errors in comparison
with the MF-PCA residuals.
}
\end{figure}

In Figure~\ref{fig:residplot_plaw}, we show the continuum residuals from $C_\textrm{mean}$,
 as a function of rest-wavelength in the \lya\ forest region. 
Comparing this plot with Figure~\ref{fig:residplot}(b), which show the MF-PCA fitting residuals
from the same $[\snr, \zq]$ bin, it is clear that
MF-PCA dramatically reduces the range of fitting errors by a factor of
3. Indeed, the $C_\textrm{mean}(\lambrest)$ residuals are significantly larger than even the worst-case
scenario for MF-PCA continua, represented in Figure~\ref{fig:residplot}(a).
The significant bias (black line) of the residuals from $C_\textrm{mean}(\lambrest)$ is
puzzling at first glance, as one would expect to recover the mean spectrum (and hence no 
bias) when averaging over large numbers of spectra.
We suspect this bias most likely due to an asymmetry in the distribution of 
 power-law spectral indices in quasars 
\citep[see][]{desjacques:2007}, which we have not accounted for in our mock spectra.  However, 
this does not affect the scatter and shape
of the continua, which is the present quantity of interest. The median RMS error from 
the continua shown in Figure~\ref{fig:residplot_plaw} is $(\delta C)_\mathrm{rms} = 0.13$, 
which is worse than any of the $[\snr, \zq]$ bins evaluated in Figure~\ref{fig:dcont_z}. 

\subsection{Residual Continuum Power}
The RMS continuum fitting error is not the most important quantity for studies
of the 1-dimensional \lya\ forest flux power spectrum, 
\begin{equation}
P_F(k) = 2 \pi \int^\infty_0 \delta(k) \delta^*(k) dk,
\end{equation}
where $\delta \equiv F/\langle F \rangle - 1$, and $k\equiv 2\pi/l$ is the Fourier wavenumber.
Rather, it is the Fourier 
power from the continuum errors which is the troublesome systematic. For example, in their
measurement of $P_F(k)$ from SDSS data, \citet{mcdonald:2006}
were limited to scales of $k \geq 0.0014\, \mathrm{km^{-1} s}$, corresponding to 
comoving distances of $r \lesssim 50 \mpc$  or $\lambrest \lesssim 18 \ang$
in the quasar restframe wavelength.
This was due to the increasing
influence of continuum power at large scales.
It is therefore pertinent to investigate the amount of residual Fourier power introduced by the
various continuum fitting methods. 

\begin{figure}
\epsscale{1.2}
\plotone{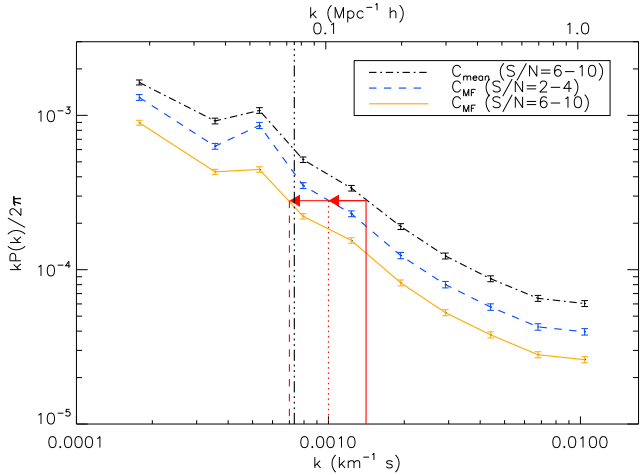}
\caption{ \label{fig:residpower}
Mean power spectrum of 1000 residuals from continuum fitting on mock spectra, 
calculated using power-law+mean continuum fitting $C_\textrm{mean}$ on $\snr=6-10$
spectra (black dot-dashed lines), and mean-flux regulated PCA fitting $C_\textrm{MF}$
on spectra with $\snr=2-4$ (yellow dashed lines) and $\snr=6-10$ (blue solid lines).
The upper abscissa shows the wave number in units of comoving distance, evaluated
at $z = 2.75$ and assuming a flat $\Lambda$CDM cosmology with $h = 0.7$ and 
$\Omega_m 0.28$. The black vertical dot-dot-dot-dashed line indicates the 
location of the first BAO peak for this cosmology. 
The error bars show the error on the mean estimated from bootstrap resampling.
The red solid line denotes the continuum-limited scale of 
$k \geq 0.0014\, \mathrm{km^{-1} s}$, the smallest $k$ at which \citet{mcdonald:2006} measured $P_F(k)$.
Red arrows indicate the points at which the $C_\textrm{MF}$ fits reach the same 
continuum-limited power as $C_\textrm{mean}$. 
The vertical red-dotted and red-dashed lines indicate the new continuum-limited scale 
for the $\snr=2-4$ and $\snr=6-10$ MF-PCA fits, respectively. 
}
\end{figure}

In Figure~\ref{fig:residpower}, we show the mean power
spectrum of the continuum residuals from 1000 mock spectra, 
$\delta C (\lambrest)$, for the power-law+mean 
continuum method, $C_\textrm{mean}$, and MF-PCA continuum fitting, $C_\textrm{MF}$, shown 
for two signal-to-noise bins. All the mock spectra had quasar redshifts 
in the range $\zq=2.9-3.1$.
All the residual power-spectra have the same overall shape, with a bump at 
$k \approx 0.0005\, \mathrm{km^{-1} s}$ corresponding to the weak emission
lines in the intrinsic quasar spectrum at scales of $\lambrest \approx 50\ang$.
This is unsurprising, since imperfections in the continuum fitting should give 
similarly-shaped residuals.
At fixed $k$, the amplitudes follow the same pattern we had already
discussed above: the residual power from MF-PCA fitting is significantly lower than that
from the power-law+mean continuum fits. Even with noisy $\snr \approx 3$ spectra, 
$C_\textrm{MF}$ continuum fitting reduces the residual power by $\sim 30\%$  
compared to $C_\textrm{mean}$ fitted to higher-\snr\ spectra. 

In their measurement of $P_F(k)$ from SDSS data, \citet{mcdonald:2006} had used a mean continuum
shape for all their spectra, which is similar to $C_\textrm{mean}$ except that they did not fit
for the individual power-law indices. 
Using this method, \citet{mcdonald:2006} found that residual continuum power started becoming problematic 
at scales greater than $k=0.0014 \, \mathrm{km^{-1} s}$, indicated by the solid red vertical line in 
Figure~\ref{fig:residpower}. 
We thus estimate the residual power
in $C_\mathrm{mean}$ at this scale at which continuum power interferes with
$P_F(k)$ measurements.
We can then look for the points in the $C_\mathrm{MF}$ residual power spectra with
the same limiting power (red arrows in Figure~\ref{fig:residpower}).
Note that this assumes that the \lya\ forest power is constant whereas the
\lya\ forest power decreases with scale, but 
the change is gradual.
For our rough estimates,  
we can approximate it as constant over small logarithmic intervals.

The corresponding limiting values of $k$ (red dotted lines) are significantly smaller than 
for $C_\mathrm{mean}$.
This suggests that MF-PCA continuum fitting could allow the \lya\ 
forest flux power spectrum to be measured at larger scales than previously possible. 
For $\snr = 6-10$ spectra, the lower $k$-limit is now $k=0.0007 \, \mathrm{km^{-1} s}$.
This corresponds to a doubling of the accessible comoving scales: 
$r=2\pi/k \approx 90 \mpc$ at $z=2.75$ compared to 
$r \approx 45 \mpc$ in the \citet{mcdonald:2006} study, where these distances are 
calculated assuming a standard flat $\Lambda$CDM cosmology with $h=0.7$, 
$\Omega_m = 0.28$ and with a $w=-1$ cosmological constant. 
Even for noisy ($\snr = 2-4$) spectra, the accessible scale
has been increased significantly to $r \approx 65 \mpc$.

The new continuum-limited scales approach the $\sim 100\mpc$  baryon acoustic oscillation 
(BAO) scale at moderate signal-to-noise ($\snr \gtrsim 6$). In Fig.~\ref{fig:residpower}, the black vertical 
dot-dot-dot-dashed line indicates the wavenumber of first BAO peak, calculated from the prescription in 
\citet{eisenstein:1998}.
Even though future BAO measurements in the 
\lya\ forest are expected to be carried out in 3-dimensions \citep{mcdonald:2007}, 
the increase in accessible modes along the lines-of-sight will improve the robustness and 
precision of the measurements.

\section{Results and Conclusion} \label{sec:results}
\subsection{Public Release of Continua}
We have carried out mean-flux regulated PCA (MF-PCA) continuum fitting on 12,069 quasar spectra from the
SDSS DR7 catalog. The continua in the spectra range $1030 \ang < \lambrest < 1600\ang$ 
have been made publicly available, and can be downloaded via anonymous 
 FTP\footnote{\url{ftp.astro.princeton.edu/lee/continua/}}.
The IDL fitting code took $\sim 0.5$ seconds per spectrum (including file input/output) 
on a single processor core of a 3.0 GHz Intel Quad Core desktop with 2 GB of RAM, 
allowing the entire SDSS DR7 \lya\ forest sample to be fitted in about two hours.  

Since we expect the MF-PCA technique to provide a good continuum fit only when there is a good PCA fit 
redwards of the quasar \lya\ line, the fitted spectra have been visually inspected to verify the fit quality
in the $\lambrest = 1216\ang - 1600 \ang$ wavelength region. 
Approximately $89\%$ of the 
spectra had reasonable PCA fits, and these have been flagged as such in the publicly available continua, 
although we recommend that users of the continua should make their own cuts on
the fit quality. 
Approximately 30\% of the spectra were better fitted by the low-redshift \citet{suzuki:2005} templates, 
while the rest were better fitted by the \citet{paris:2011} templates. 
This is qualitatively as expected, since there is greater overlap by the \citet{paris:2011} templates 
in the luminosity distribution of the SDSS quasars.

From the mock spectra analysis of the fit quality in \S~\ref{sec:mocks}, we have estimated the continuum fitting error 
at each pixel within the \lya\ forest, as a function of quasar redshift and spectral signal-to-noise.
However, the errors have significant covariances, therefore it would be too unwieldy to provide 
the full error estimates for each spectrum although we can provide them upon request.

It is worth noting that although we have made a choice on the mean-flux of the \lya\ 
forest in our fits (Eq~\ref{eq:fmeanz}), it is straightforward for users to rescale each fitted continuum
 to their favorite $\langle F \rangle (z)$. 

\subsection{Empirical Tests of Fit Quality}

While we have studied the performance of MF-PCA continuum fitting on mock spectra in \S~\ref{sec:mocks},
it is difficult to empirically constrain the quality of the fits. 
One possibility is to compare a small subset of the data with high-resolution, high-\snr\ spectra 
of the same objects. 
However, even with high-resolution spectra, it is questionable whether there are sufficient
transmission peaks in the forest to adequately constrain the continuum shape; \citet{fg+08} have shown
 that accurate fitting of the quasar continuum is difficult beyond $z \approx 2.5$. 
 Furthermore, most high-resolution spectra are obtained from echelle spectrographs
 with uncertain spectrophotometry, so it would be tricky to directly compare quasar sightlines
 which have been observed in both SDSS and high-resolution echelle spectrographs.

However, it is possible to get a sense of the efficacy of our MF-PCA continua by 
stacking large numbers of spectra. 
This cancels out the \lya\ forest power from individual sightlines, 
and allows the underlying continuum shape to be seen, albeit lowered due to the mean \lya\ absorption.

\begin{figure*}
\plotone{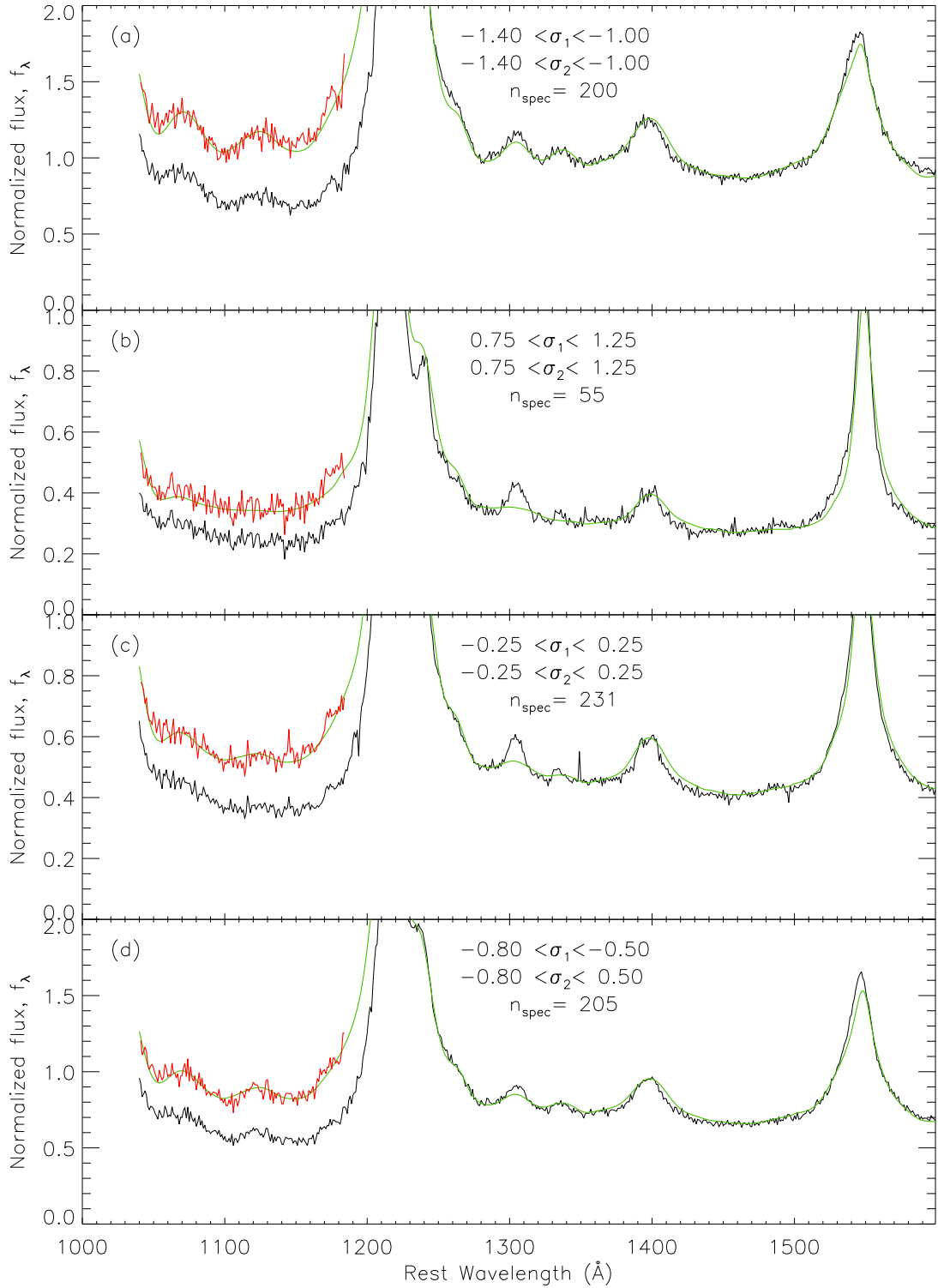}
\caption{ \label{fig:qsoclass}
Stacked SDSS \lya\ forest spectra (black) and similarly-stacked
MF-PCA continuum fits, plotted for narrow selections of the 
first two PCA eigenvalues, $\sigma_1$ and $\sigma_2$.
The red curve shows the $1041 \ang < \lambrest < 1185$ \lya\ forest region
of the spectra which have been corrected by the mean-flux prior to stacking.
The agreement of the MF-PCA stacks with the mean-flux corrected \lya\
forest stacks show that the MF-PCA is doing a good job of predicting the
shape of the \lya\ forest continuum.
}
\end{figure*}

Recall that the PCA coefficients, $c_j$, parametrize the shape of the quasar spectra.
 Therefore, if our continuum-fitting technique works, quasars with similar $c_j$ measured from 
 $\lambrest > 1216\ang$ should have similar-looking continua within the \lya\ forest region. 
Thanks to the large number of spectra in our SDSS sample, it is possible to stack 
$\sim 10^2$ spectra with similar values of $c_j$ to recover the collective shape
of their \lya\ forest continua. 

We can select subsamples of quasars based on their values of 
$\sigma_1 \equiv c_1/ \lambda_1 $ and $\sigma_2 \equiv c_2/ \lambda_2$, where
$\lambda_1=7.563$ and $\lambda_2=3.604$ are the standard deviations of $c_1$ and $c_2$, 
respectively, in the low-redshift HST eigenspectra \citep{suzuki:2006}.
These two principal components account for approximately 80\% of the total variance in the low-redshift
quasar templates. We limit ourselves to spectra with $\snr > 3 \, \text{pixel}^{-1}$, and which have
been visually inspected to be decent fits redwards of $\lambrest=1216\ang$.
We also select quasars with $\zq > 2.6$ in order to ensure reasonably complete
coverage of the \lya\ forest.
Within a subsample, each spectrum is first normalized near $\lambrest = 1280\ang$ and 
rebinned into a common wavelength grid with $\Delta \lambrest = 1\ang$ bins before being stacked. 
The same procedure is carried out on the MF-PCA continuum fitted to each spectrum,
to obtain a mean MF-PCA continuum for the subsample. 

In Fig.~\ref{fig:qsoclass} we show 4 subsamples from our SDSS sample with different 
$[\sigma_1, \sigma_2]$ with respect to the low-redshift \citet{suzuki:2005} eigenspectra.
Redwards of $1216\ang$, we see that the least-squared PCA procedure generally 
does a good job of fitting the emission lines, although
there are inaccuracies in fitting \waveion{N}{5}{1240} and \waveion{Si}{2}{1306}.
Bluewards of $1216\ang$, the stacked spectrum appears to have a similar shape to the 
fitted continua, although the overall flux level is depressed due to the mean \lya\ absorption. 

We can make a more direct comparison between the stacked spectra and the fitted \lya\ forest continua by 
correcting each observed \lya\ forest pixel by
its mean flux (using Equation~\ref{eq:fmeanz}) before stacking. 
The mean-flux corrected \lya\ forest is shown as the disembodied red line in Fig.~\ref{fig:qsoclass}.
It is gratifying to see that the stacked MF-PCA continua generally agrees well with the stacked 
\lya\ forest spectra. Our technique can clearly account for the diversity in quasar continua: 
spectra with clear emission-line features  (Fig~\ref{fig:qsoclass}a) and those with smooth continua
(Fig~\ref{fig:qsoclass}b) are well differentiated. Because we have corrected each \lya\ forest
pixel by the same mean-flux (Eq.~\ref{eq:fmeanz}) which we have used to carry out the MF-PCA
fits, we expect the amplitude of the corrected \lya\ forest stacks, 
in $\lambrest = 1041\ang-1185\ang$, 
to match those of the stacked MF-PCA continua, but the tilt and shape of the continuum bears
testament to the success of the technique.
In addition, since the MF-PCA continua shown in Fig~\ref{fig:qsoclass} were a subset which used the
\citet{suzuki:2005} quasar templates, this suggests that it was appropriate to use
low-redshift templates to fit some of the $\zq \gtrsim 2$ SDSS spectra.

\subsection{Conclusions}

We have introduced mean-flux regulated PCA (MF-PCA) continuum fitting, 
a new technique for predicting the \lya\ forest continuum in low-\snr\ spectra. 
In tests on mock spectra, we have
found that MF-PCA can predict the continuum at the $8\%$ RMS level in SDSS spectra with 
$\snr \sim 2$ at $z=2.5$, and $<5\%$ RMS in $\snr \gtrsim 5$ spectra. 
This is a significant improvement over the $\sim 15\%$ RMS continuum errors previously
achievable in low-\snr\ spectra.
We are making available MF-PCA continuum fits for 12,069 \lya\ forest spectra from the SDSS 
DR7 quasar catalog.
The MF-PCA technique also significantly reduces the Fourier power from continuum-fitting residuals
by a factor of a few in comparison with dividing by a mean continuum. 
This will allow a concomitant increase in the accessible scales for \lya\ forest
flux power spectrum measurements. 

This improved continuum-fitting accuracy will significantly increase the value of low-\snr\ 
\lya\ forest data. For example, the ongoing Baryon Oscillations Spectroscopic Survey (BOSS)
will obtain \lya\ forest spectra from $\sim 150,000$ quasars at $\zq \gtrsim 2$, 
with the aim of measuring the baryon acoustic oscillation feature in the \lya\ forest absorption 
across different quasar sightlines.
The typical signal-to-noise ($\snr \sim 2$) of BOSS spectra will be even 
lower than that of SDSS ($\snr \sim 4$), therefore we expect the MF-PCA technique to
contribute significantly to the utility of the BOSS data.

There are several ways in which the current work could be improved. 
The \citet{suzuki:2005} PCA templates with which we have used to fit 
some of the spectra were derived from a low-redshift quasar sample which may not be 
a perfect descriptor of the SDSS data (although the test in \S~\ref{sec:results} shows that
it does a reasonable job). In addition, while the $\zq \sim 3$ \citet{paris:2011} templates were indeed 
obtained from SDSS quasars, they used a high-luminosity subset which are not representative
of the full SDSS luminosity distribution. Furthermore, the hand-fitting technique which they had
used to obtain continua from these spectra cannot be used for the lower-luminosity 
(and hence lower-\snr) quasars.

However, with a large data set such as SDSS or BOSS, it is possible to regard the \lya\ forest
absorption within individual spectra as a noise term which cancels out with 
sufficiently large numbers of template spectra. 
This will allow new eigenspectra to be generated from the data itself,
although each individual spectrum will need to be corrected by its mean flux
before being included in the eigenspectrum solution. 
In the near future, we will work on this technique to generate new eigenspectra from the BOSS data.

The other issue with the MF-PCA fitting is that it requires an assumed mean-flux, 
$\langle F \rangle (z)$ for the \lya\ forest. This is not ideal, as the evolution of the 
mean-flux is an important observable of the \lya\ forest. 
This could in principle be overcome by solving simultaneously for the mean-flux of the \lya\ forest
and the continuum-fitting parameters for the individual spectra, using maximum-likelihood
techniques. 
This would allow large \lya\ forest data sets to be continuum-fitted and studied in a fully self-consistent fashion.

\acknowledgements{
The authors thank Michael Strauss and Xavier Prochaska for useful
discussions and comments, and to Is\^{a}belle Paris for providing 
the data from her mean-flux measurement.

    Funding for the SDSS and SDSS-II has been provided by the Alfred P. Sloan Foundation, the Participating Institutions, the National Science Foundation, the U.S. Department of Energy, the National Aeronautics and Space Administration, the Japanese Monbukagakusho, the Max Planck Society, and the Higher Education Funding Council for England. The SDSS Web Site is 
    \url{http://www.sdss.org/}.

    The SDSS is managed by the Astrophysical Research Consortium for the Participating Institutions: the American Museum of Natural History, Astrophysical Institute Potsdam, University of Basel, University of Cambridge, Case Western Reserve University, University of Chicago, Drexel University, Fermilab, the Institute for Advanced Study, the Japan Participation Group, Johns Hopkins University, the Joint Institute for Nuclear Astrophysics, the Kavli Institute for Particle Astrophysics and Cosmology, the Korean Scientist Group, the Chinese Academy of Sciences (LAMOST), Los Alamos National Laboratory, the Max-Planck-Institute for Astronomy (MPIA), the Max-Planck-Institute for Astrophysics (MPA), New Mexico State University, Ohio State University, University of Pittsburgh, University of Portsmouth, Princeton University, the United States Naval Observatory, and the University of Washington.  \\ \hfill. }

\bibliographystyle{apj}
\bibliography{ms,apj-jour}

\begin{thebibliography}{29}
\expandafter\ifx\csname natexlab\endcsname\relax\def\natexlab#1{#1}\fi

\bibitem[{{Allen} {et~al.}(2011){Allen}, {Hewett}, {Maddox}, {Richards}, \&
  {Belokurov}}]{allen:2011}
{Allen}, J.~T., {Hewett}, P.~C., {Maddox}, N., {Richards}, G.~T., \&
  {Belokurov}, V. 2011, \mnras, 410, 860

\bibitem[{{Baldwin} {et~al.}(1978){Baldwin}, {Burke}, {Gaskell}, \&
  {Wampler}}]{baldwin:1978}
{Baldwin}, J.~A., {Burke}, W.~L., {Gaskell}, C.~M., \& {Wampler}, E.~J. 1978,
  \nat, 273, 431

\bibitem[{{Bernardi} {et~al.}(2003){Bernardi}, {Sheth}, {SubbaRao}, {Richards},
  {Burles}, {Connolly}, {Frieman}, {Nichol}, {Schaye}, {Schneider}, {Vanden
  Berk}, {York}, {Brinkmann}, \& {Lamb}}]{bernardi:2003}
{Bernardi}, M., {Sheth}, R.~K., {SubbaRao}, M., {Richards}, G.~T., {Burles},  S., {Connolly}, A.~J., {Frieman}, J., {Nichol}, R., {et al.} 2003, \aj, 125, 32

\bibitem[{{Dall'Aglio} {et~al.}(2009){Dall'Aglio}, {Wisotzki}, \&
  {Worseck}}]{dall+09}
{Dall'Aglio}, A., {Wisotzki}, L., \& {Worseck}, G. 2009, ArXiv e-prints

\bibitem[{{Desjacques} {et~al.}(2007){Desjacques}, {Nusser}, \&
  {Sheth}}]{desjacques:2007}
{Desjacques}, V., {Nusser}, A., \& {Sheth}, R.~K. 2007, \mnras, 374, 206

\bibitem[{{Draine}(2011)}]{draine:2011}
{Draine}, B.~T. 2011, {Physics of the Interstellar and Intergalactic Medium},
  ed. {Draine, B.~T.} (Princeton University Press)

\bibitem[{{Eisenstein} \& {Hu}(1998)}]{eisenstein:1998}
{Eisenstein}, D.~J. \& {Hu}, W. 1998, \apj, 496, 605

\bibitem[{{Fan}(2006)}]{fan:2006}
{Fan}, X. 2006, New Astronomy Reviews, 50, 665

\bibitem[{{Faucher-Gigu{\`e}re} {et~al.}(2008){Faucher-Gigu{\`e}re},
  {Prochaska}, {Lidz}, {Hernquist}, \& {Zaldarriaga}}]{fg+08}
{Faucher-Gigu{\`e}re}, C., {Prochaska}, J.~X., {Lidz}, A., {Hernquist}, L., \&
  {Zaldarriaga}, M. 2008, \apj, 681, 831

\bibitem[{{Francis} {et~al.}(1992){Francis}, {Hewett}, {Foltz}, \&
  {Chaffee}}]{francis:1992}
{Francis}, P.~J., {Hewett}, P.~C., {Foltz}, C.~B., \& {Chaffee}, F.~H. 1992,
  \apj, 398, 476

\bibitem[{{Hewett} \& {Wild}(2010)}]{hewett:2010}
{Hewett}, P.~C. \& {Wild}, V. 2010, \mnras, 405, 2302

\bibitem[{{Lee}(2011)}]{lee:2011}
{Lee}, K.-G. 2011, ArXiv e-prints

\bibitem[{{Lee} \& {Spergel}(2011)}]{lee:2011a}
{Lee}, K.-G. \& {Spergel}, D.~N. 2011, \apj, 734, 21

\bibitem[{{Markwardt}(2009)}]{markwardt:2009}
{Markwardt}, C.~B. 2009, in Astronomical Society of the Pacific Conference
  Series, Vol. 411, Astronomical Data Analysis Software and Systems XVIII, ed.
  {D.~A.~Bohlender, D.~Durand, \& P.~Dowler}, 251--+

\bibitem[{{McDonald} \& {Eisenstein}(2007)}]{mcdonald:2007}
{McDonald}, P. \& {Eisenstein}, D.~J. 2007, \prd, 76, 063009

\bibitem[{{McDonald} {et~al.}(2006){McDonald}, {Seljak}, {Burles}, {Schlegel},
  {Weinberg}, {Cen}, {Shih}, {Schaye}, {Schneider}, {Bahcall}, {Briggs},
  {Brinkmann}, {Brunner}, {Fukugita}, {Gunn}, {Ivezi{\'c}}, {Kent}, {Lupton},
  \& {Vanden Berk}}]{mcdonald:2006}
{McDonald}, P., {Seljak}, U., {Burles}, S., {Schlegel}, D.~J., {Weinberg},  D.~H., {Cen}, R., {Shih}, D., {Schaye}, J., {et al.} 2006, \apjs, 163, 80

\bibitem[{{Noterdaeme} {et~al.}(2009){Noterdaeme}, {Petitjean}, {Ledoux}, \&
  {Srianand}}]{noterdaeme:2009}
{Noterdaeme}, P., {Petitjean}, P., {Ledoux}, C., \& {Srianand}, R. 2009, \aap,
  505, 1087

\bibitem[{{P{\^a}ris} {et~al.}(2011){P{\^a}ris}, {Petitjean}, {Rollinde},
  {Aubourg}, {Busca}, {Charlassier}, {Delubac}, {Hamilton}, {Le Goff},
  {Palanque-Delabrouille}, {Peirani}, {Pichon}, {Rich}, {Vargas-Maga{\~n}a}, \&
  {Y{\`e}che}}]{paris:2011}
{P{\^a}ris}, I., {Petitjean}, P., {Rollinde}, E., {Aubourg}, E., {Busca}, N.,  {Charlassier}, R., {Delubac}, T., {Hamilton}, J.-C., {et al.} 2011, \aap, 530, A50+

\bibitem[{{Schneider} {et~al.}(2010){Schneider}, {Richards}, {Hall}, {Strauss},
  {Anderson}, {Boroson}, {Ross}, {Shen}, {Brandt}, {Fan}, {Inada}, {Jester},
  {Knapp}, {Krawczyk}, {Thakar}, {Vanden Berk}, {Voges}, {Yanny}, {York},
  {Bahcall}, {Bizyaev}, {Blanton}, {Brewington}, {Brinkmann}, {Eisenstein},
  {Frieman}, {Fukugita}, {Gray}, {Gunn}, {Hibon}, {Ivezi{\'c}}, {Kent}, {Kron},
  {Lee}, {Lupton}, {Malanushenko}, {Malanushenko}, {Oravetz}, {Pan}, {Pier},
  {Price}, {Saxe}, {Schlegel}, {Simmons}, {Snedden}, {SubbaRao}, {Szalay}, \&
  {Weinberg}}]{schneider:2010}
{Schneider}, D.~P., {Richards}, G.~T., {Hall}, P.~B., {Strauss}, M.~A.,  {Anderson}, S.~F., {Boroson}, T.~A., {Ross}, N.~P., {Shen}, Y., {et al.} 2010, \aj, 139, 2360

\bibitem[{{Shen} {et~al.}(2011){Shen}, {Richards}, {Strauss}, {Hall},
  {Schneider}, {Snedden}, {Bizyaev}, {Brewington}, {Malanushenko},
  {Malanushenko}, {Oravetz}, {Pan}, \& {Simmons}}]{shen:2011}
{Shen}, Y., {Richards}, G.~T., {Strauss}, M.~A., {Hall}, P.~B., {Schneider},  D.~P., {Snedden}, S., {Bizyaev}, D., {Brewington}, H., {et al.} 2011, \apjs,  194, 45

\bibitem[{{Slosar} {et~al.}(2011){Slosar}, {Font-Ribera}, {Pieri}, {Rich}, {Le
  Goff}, {Aubourg}, {Brinkmann}, {Busca}, {Carithers}, {Charlassier},
  {Cort{\^e}s}, {Croft}, {Dawson}, {Eisenstein}, {Hamilton}, {Ho}, {Lee},
  {Lupton}, {McDonald}, {Medolin}, {Miralda-Escud{\'e}}, {Muna}, {Myers},
  {Nichol}, {Palanque-Delabrouille}, {P{\^a}ris}, {Petitjean}, {Pi{\v s}kur},
  {Rollinde}, {Ross}, {Schlegel}, {Schneider}, {Sheldon}, {Weaver}, {Weinberg},
  {Yeche}, \& {York}}]{slosar:2011}
{Slosar}, A., {Font-Ribera}, A., {Pieri}, M.~M., {Rich}, J., {Le Goff}, J.-M.,  {Aubourg}, {\'E}., {Brinkmann}, J., {Busca}, N., {et al.} 2011, ArXiv e-prints

\bibitem[{{Stoughton} {et~al.}(2002){Stoughton}, {Lupton}, {Bernardi},
  {Blanton}, {Burles}, {Castander}, {Connolly}, {Eisenstein}, {Frieman},
  {Hennessy}, {Hindsley}, {Ivezi{\'c}}, {Kent}, {Kunszt}, {Lee}, {Meiksin},
  {Munn}, {Newberg}, {Nichol}, {Nicinski}, {Pier}, {Richards}, {Richmond},
  {Schlegel}, {Smith}, {Strauss}, {SubbaRao}, {Szalay}, {Thakar}, {Tucker},
  {Vanden Berk}, {Yanny}, {Adelman}, {Anderson}, {Anderson}, {Annis},
  {Bahcall}, {Bakken}, {Bartelmann}, {Bastian}, {Bauer}, {Berman},
  {B{\"o}hringer}, {Boroski}, {Bracker}, {Briegel}, {Briggs}, {Brinkmann},
  {Brunner}, {Carey}, {Carr}, {Chen}, {Christian}, {Colestock}, {Crocker},
  {Csabai}, {Czarapata}, {Dalcanton}, {Davidsen}, {Davis}, {Dehnen},
  {Dodelson}, {Doi}, {Dombeck}, {Donahue}, {Ellman}, {Elms}, {Evans}, {Eyer},
  {Fan}, {Federwitz}, {Friedman}, {Fukugita}, {Gal}, {Gillespie}, {Glazebrook},
  {Gray}, {Grebel}, {Greenawalt}, {Greene}, {Gunn}, {de Haas}, {Haiman},
  {Haldeman}, {Hall}, {Hamabe}, {Hansen}, {Harris}, {Harris}, {Harvanek},
  {Hawley}, {Hayes}, {Heckman}, {Helmi}, {Henden}, {Hogan}, {Hogg}, {Holmgren},
  {Holtzman}, {Huang}, {Hull}, {Ichikawa}, {Ichikawa}, {Johnston}, {Kauffmann},
  {Kim}, {Kimball}, {Kinney}, {Klaene}, {Kleinman}, {Klypin}, {Knapp},
  {Korienek}, {Krolik}, {Kron}, {Krzesi{\'n}ski}, {Lamb}, {Leger},
  {Limmongkol}, {Lindenmeyer}, {Long}, {Loomis}, {Loveday}, {MacKinnon},
  {Mannery}, {Mantsch}, {Margon}, {McGehee}, {McKay}, {McLean}, {Menou},
  {Merelli}, {Mo}, {Monet}, {Nakamura}, {Narayanan}, {Nash}, {Neilsen},
  {Newman}, {Nitta}, {Odenkirchen}, {Okada}, {Okamura}, {Ostriker}, {Owen},
  {Pauls}, {Peoples}, {Peterson}, {Petravick}, {Pope}, {Pordes}, {Postman},
  {Prosapio}, {Quinn}, {Rechenmacher}, {Rivetta}, {Rix}, {Rockosi}, {Rosner},
  {Ruthmansdorfer}, {Sandford}, {Schneider}, {Scranton}, {Sekiguchi}, {Sergey},
  {Sheth}, {Shimasaku}, {Smee}, {Snedden}, {Stebbins}, {Stubbs}, {Szapudi},
  {Szkody}, {Szokoly}, {Tabachnik}, {Tsvetanov}, {Uomoto}, {Vogeley}, {Voges},
  {Waddell}, {Walterbos}, {Wang}, {Watanabe}, {Weinberg}, {White}, {White},
  {Wilhite}, {Wolfe}, {Yasuda}, {York}, {Zehavi}, \& {Zheng}}]{stoughton:2002}
{Stoughton}, C., {Lupton}, R.~H., {Bernardi}, M., {Blanton}, M.~R., {Burles},  S., {Castander}, F.~J., {Connolly}, A.~J., {Eisenstein}, D.~J., {et al.} 2002,  \aj, 123, 485

\bibitem[{{Suzuki}(2006)}]{suzuki:2006}
{Suzuki}, N. 2006, \apjs, 163, 110

\bibitem[{{Suzuki} {et~al.}(2005){Suzuki}, {Tytler}, {Kirkman}, {O'Meara}, \&
  {Lubin}}]{suzuki:2005}
{Suzuki}, N., {Tytler}, D., {Kirkman}, D., {O'Meara}, J.~M., \& {Lubin}, D.
  2005, \apj, 618, 592

\bibitem[{{Telfer} {et~al.}(2002){Telfer}, {Zheng}, {Kriss}, \&
  {Davidsen}}]{telfer:2002}
{Telfer}, R.~C., {Zheng}, W., {Kriss}, G.~A., \& {Davidsen}, A.~F. 2002, \apj,
  565, 773

\bibitem[{{Tytler} {et~al.}(2004){Tytler}, {Kirkman}, {O'Meara}, {Suzuki},
  {Orin}, {Lubin}, {Paschos}, {Jena}, {Lin}, {Norman}, \&
  {Meiksin}}]{tytler:2004}
{Tytler}, D., {Kirkman}, D., {O'Meara}, J.~M., {Suzuki}, N., {Orin}, A.,
  {Lubin}, D., {Paschos}, P., {Jena}, T., {Lin}, W., {Norman}, M.~L., \&
  {Meiksin}, A. 2004, \apj, 617, 1

\bibitem[{{vanden Berk} {et~al.}(2004){vanden Berk}, {Yip}, {Connolly},
  {Jester}, \& {Stoughton}}]{vanden-berk:2004}
{vanden Berk}, D., {Yip}, C., {Connolly}, A., {Jester}, S., \& {Stoughton}, C.
  2004, in Astronomical Society of the Pacific Conference Series, Vol. 311, AGN
  Physics with the Sloan Digital Sky Survey, ed. {G.~T.~Richards \&
  P.~B.~Hall}, 21--+

\bibitem[{{Vanden Berk} {et~al.}(2001){Vanden Berk}, {Richards}, {Bauer},
  {Strauss}, {Schneider}, {Heckman}, {York}, {Hall}, {Fan}, {Knapp},
  {Anderson}, {Annis}, {Bahcall}, {Bernardi}, {Briggs}, {Brinkmann}, {Brunner},
  {Burles}, {Carey}, {Castander}, {Connolly}, {Crocker}, {Csabai}, {Doi},
  {Finkbeiner}, {Friedman}, {Frieman}, {Fukugita}, {Gunn}, {Hennessy},
  {Ivezi{\'c}}, {Kent}, {Kunszt}, {Lamb}, {Leger}, {Long}, {Loveday}, {Lupton},
  {Meiksin}, {Merelli}, {Munn}, {Newberg}, {Newcomb}, {Nichol}, {Owen}, {Pier},
  {Pope}, {Rockosi}, {Schlegel}, {Siegmund}, {Smee}, {Snir}, {Stoughton},
  {Stubbs}, {SubbaRao}, {Szalay}, {Szokoly}, {Tremonti}, {Uomoto}, {Waddell},
  {Yanny}, \& {Zheng}}]{vanden-berk:2001}
{Vanden Berk}, D.~E., {Richards}, G.~T., {Bauer}, A., {Strauss}, M.~A.,  {Schneider}, D.~P., {Heckman}, T.~M., {York}, D.~G., {Hall}, P.~B., {et al.} 2001, \aj, 122, 549

\bibitem[{{White} {et~al.}(2010){White}, {Pope}, {Carlson}, {Heitmann},
  {Habib}, {Fasel}, {Daniel}, \& {Lukic}}]{white+10}
{White}, M., {Pope}, A., {Carlson}, J., {Heitmann}, K., {Habib}, S., {Fasel},
  P., {Daniel}, D., \& {Lukic}, Z. 2010, \apj, 713, 383

\bibitem[{{Zheng} {et~al.}(1997){Zheng}, {Kriss}, {Telfer}, {Grimes}, \&
  {Davidsen}}]{zheng:1997}
{Zheng}, W., {Kriss}, G.~A., {Telfer}, R.~C., {Grimes}, J.~P., \& {Davidsen},
  A.~F. 1997, \apj, 475, 469

\end{thebibliography}
\end{document}